\documentclass[11pt, reqno]{amsart}
\usepackage{graphicx,amsmath,amssymb,epsfig,xcolor,enumerate}
\usepackage{booktabs}
\usepackage{amsfonts,bbm}
\usepackage{harvard}
\usepackage{epstopdf}
\usepackage{afterpage}
\usepackage{graphicx}
\usepackage{float,caption}
\usepackage{subcaption}
\usepackage{hypernat}
\definecolor{magenta(dye)}{rgb}{0.79, 0.08, 0.48}
\usepackage[
  bookmarks=true, 
  bookmarksopen=true, 
  breaklinks=true, 
  colorlinks=true,
  linkcolor=magenta(dye),
  citecolor=magenta(dye), 
]{hyperref}
\usepackage{mathrsfs}

\usepackage{hyperref}

\hypersetup{
    colorlinks=true,
    urlcolor=blue,
    linkcolor=blue,
    citecolor=blue
}

\newtheorem{theorem}{Theorem}
\theoremstyle{plain}

\newtheorem{definition}{Definition}
\newtheorem{example}{Example}

\newtheorem{proposition}{Proposition}

\newtheorem{assumption}{Assumption}

\input{tcilatex}
\graphicspath{{./QE/illustrations/}}

\begin{document}
\title{Duality in dynamic discrete choice models}
\author[Chiong]{Khai X. Chiong{\small $^{\S }$}}
\author[Galichon]{Alfred Galichon\textit{$^{\dag }$}}
\author[Shum]{Matt Shum$^{\clubsuit }$}
\date{First draft: April 2013. This version: May 2015.\\
The authors thank the Editor, three anonymous referees, as well as Benjamin
Connault, Thierry Magnac, Emerson Melo, Bob Miller, Sergio Montero, John
Rust, Sorawoot (Tang) Srisuma, and Haiqing Xu for useful comments. We are
especially grateful to Guillaume Carlier for providing decisive help with
the proof of Theorem~\ref{thm:consistency}. We also thank audiences at
Michigan, Northwestern, NYU, Pittsburgh, UCSD, the CEMMAP conference on inference in
game-theoretic models (June 2013), UCLA econometrics mini-conference (June
2013), the Boston College Econometrics of Demand Conference (December 2013)
and the Toulouse conference on \textquotedblleft Recent Advances in Set
Identification\textquotedblright\ (December 2013) for helpful comments.
Galichon's research has received funding from the European Research Council
under the European Union's Seventh Framework Programme (FP7/2007-2013) / ERC
grant agreement n%
${{}^\circ}$%
313699 and from FiME, Laboratoire de Finance des March\'{e}s de l'Energie
(www.fime-lab.org). \\
{\indent \Small $^{\S }$Division of the Humanities and Social Sciences,
California Institute of Technology; kchiong@caltech.edu }\\
\indent $^{\dag }$Department of Economics, Sciences Po;
alfred.galichon@sciences-po.fr \\
\indent $^{\clubsuit }$Division of the Humanities and Social Sciences,
California Institute of Technology; mshum@caltech.edu}

\begin{abstract}
Using results from convex analysis, we investigate a novel approach to
identification and estimation of discrete choice models which we call the
\textquotedblleft Mass Transport Approach\textquotedblright\ (MTA). We show
that the conditional choice probabilities and the choice-specific payoffs in
these models are related in the sense of \emph{conjugate duality}, and that
the identification problem is a mass transport problem. Based on this, we
propose a new two-step estimator for these models; interestingly, the first
step of our estimator involves solving a linear program which is identical
to the classic assignment (two-sided matching) game of Shapley and Shubik
(1971). The application of convex-analytic tools to dynamic discrete choice
models, and the connection with two-sided matching models, is new in the
literature. 
\end{abstract}

\maketitle

\newpage

\pagenumbering{gobble}

\linespread{1.5}

\section{{\protect\normalsize \setcounter{page}{2}\setcounter{equation}{0}%
Introduction}}

{ Empirical research utilizing dynamic discrete choice models of
economic decision-making has flourished in recent decades, with applications
in all areas of applied microeconomics including labor economics, industrial
organization, public finance, and health economics. The existing literature
on the identification and estimation of these models has recognized a close
link between the conditional choice probabilities (hereafter, CCP, which can
be observed and estimated from the data) and the payoffs (or\emph{\
choice-specific value functions}, which are unobservable to the researcher);
indeed, most estimation procedures contain an \textquotedblleft
inversion\textquotedblright\ step in which the choice-specific value
functions are recovered given the estimated choice probabilities. }

{ This paper has two contributions. First, we explicitly
characterize this duality relationship between the choice probabilities and
choice-specific payoffs. Specifically, in discrete choice models, the social
surplus function (McFadden (1978)) provides us with the mapping from payoffs
to the probabilities with which a choice is chosen at each state
(conditional choice probabilities). Recognizing that the social surplus
function is convex, we develop the idea that the convex conjugate of the
social surplus function gives us the inverse mapping - from choice
probabilities to utility indices. More precisely, the subdifferential of the
convex conjugate is a correspondence that maps from the observed choice
probabilities to an identified set of payoffs. In short, the choice
probabilities and utility indices are related in the sense of \emph{%
conjugate duality}. The discovery of this relationship allows us to
succinctly characterize the empirical content of discrete choice models,
both static and dynamic. }

{ 
}

{ Not only is the convex conjugate of the social surplus function
a useful theoretical object; it also provides a new and practical way to
\textquotedblleft invert\textquotedblright\ from a given vector of choice
probabilities back to the underlying utility indices which generated these
probabilities. This is the second contribution of this paper. We show how
the conjugate along with its set of subgradients can be efficiently computed
by means of linear programming. This linear programming formulation has the
structure of an optimal assignment problem (as in Shapley-Shubik's (1971)
classic work). This surprising connection enables us to apply insights
developed in the optimal transport literature, e.g. Villani (2003, 2009), to
discrete choice models. We call this new methodology the \textquotedblleft
Mass Transport Approach\textquotedblright\ to CCP\ inversion. }

{ This paper focuses on the estimation of dynamic discrete-choice
models via two-step estimation procedures in which conditional choice
probabilities are estimated in the initial stage; this estimation approach
was pioneered in Hotz and Miller (HM, 1993) and Hotz, Miller, Sanders, Smith
(1994).\footnote{{ Subsequent contributions include
Aguirregabiria and Mira (2002, 2007), Magnac and Thesmar (2002), Pesendorfer
and Schmidt-Dengler (2008), Bajari, et. al. (2009), Arcidiacono and Miller
(2011), and Norets and Tang (2013).}} Our use of tools and concepts from
convex analysis to study identification and estimation in this dynamic
discrete choice setting is novel in the literature. Based on our findings,
we propose a new two-step estimator for DDC models. A nice feature of our
estimator is that it works for practically any assumed distribution of the
utility shocks.\footnote{{ While existing identification results
for dynamic discrete choice models allow for quite general specifications of
the additive choice-specific utility shocks, many applications of these
two-step estimators maintain the restrictive assumption that the utility
shocks are distributed i.i.d. type I extreme value, independently of the
state variables, leading to choice probabilities which take the multinomial
logit form.}} Thus, our estimator would make possible the task of evaluating
the robustness of estimation to different distributional assumptions.%
\footnote{{ While they are not the focus in
this paper, many applications of dynamic choice models do not utilize
HM-type two step estimation procedures, and they allow for quite flexible
distributions of the utility shocks, and also for serial correlation in
these shocks (examples include Pakes (1986) and Keane and Wolpin (1997)).
This literature typically employs simulated method of moments, or simulated
maximum likelihood for estimation (see Rust (1994, section 3.3)).}} }

{ 
}

{ 
}

{ 
}

{ Section 2 contains our main results regarding duality between
choice probabilities and payoffs in discrete choice models. Based on these
results, we propose, in Section 3, a two-step estimation approach for these
models. We also emphasize here the surprising connection between dynamic
discrete-choice and optimal matching models. In Section 4 we discuss
computational details for our estimator, focusing on the use of linear
programming to compute (approximately) the convex conjugate function from
the dynamic discrete-choice model. Monte Carlo experiments (in Section 5)
show that our estimator performs well in practice, and we apply the
estimator to Rust's (1987) bus engine replacement data (Section 6). Section
7 concludes. The Appendix contains proofs and also a brief primer on
relevant results from convex analysis.  Sections 2.2 and 2.3, as
well as Section 4, are not specific to dynamic discrete choice problems but
are also true for any (static) discrete choice model. }

\section{{\protect\normalsize Basic Model\label{sec:Basic}}}

\subsection{{\protect\normalsize The framework\label{par:framework}}}

{ In this section we review the basic dynamic discrete-choice
setup, as encapsulated in Rust's (1987) seminal paper. The state variable is 
$x\in \mathcal{X}$ which we assume to take only a finite number of values.
Agents choose actions $y\in \mathcal{Y}$ from a finite space $\mathcal{Y}%
=\left\{0,1,\ldots,D\right\}$. The single-period utility flow which an agent
derives from choosing $y$ in a given period is 
\begin{equation*}
\bar{u}_{y}\left( x\right) +\varepsilon _{y}
\end{equation*}%
where $\varepsilon _{y}$ denotes the utility shock pertaining to action $y$,
which differs across agents. Across agents and time periods, the set of
utility shocks $\varepsilon \equiv \left( \varepsilon _{y}\right) _{y\in 
\mathcal{Y}}$ is distributed according to a joint distribution function $%
Q(\cdots ;x)$ which can depend on the current values of the state variable $%
x $. We assume that this distribution $Q$ is known to the researcher. }

{ Throughout, we consider a stationary setting in which the
agent's decision environment remains unchanged across time periods; thus,
for any given period, we use primes ($^{\prime}$) to denote next-period
values. Following Rust (1987), and most of the subsequent papers in this
literature, we maintain the following conditional independence assumption
(which rules out serially persistent forms of unobserved heterogeneity%
\footnote{{ See Norets (2009), Kasahara and Shimotsu (2009),
Arcidiacono and Miller (2011), and Hu and Shum (2012).}}): }

\begin{assumption}[Conditional Independence]
{ \label{ass:conditionalIndependence} $(x,\varepsilon)$ evolves
across time periods as a controlled first-order Markov process, with
transition 
\begin{align*}
{Pr}(x^{\prime},\varepsilon^{\prime}|y,x,\varepsilon )=& {Pr}(\varepsilon
^{\prime}|x^{\prime},y,x,\varepsilon )\cdot {Pr}(x^{\prime}|y,x,\varepsilon )
\\
=& {Pr}(\varepsilon^{\prime}|x^{\prime})\cdot {Pr}(x^{\prime}|y,x).
\end{align*}
}
\end{assumption}

{ The discount rate is $\beta$. Agents are dynamic optimizers
whose choices each period satisfy\footnote{{ We have used
Assumption 1 to eliminate $\varepsilon$ as a conditioning variable in the
expectation in Eq. (\ref{RandomAlternative}). }} 
\begin{equation}
y\in \arg \max_{\tilde{y}\in \mathcal{Y}}\left\{ \bar{u}_{\tilde{y}}\left( x\right) +\varepsilon _{\tilde{y}}+\beta \mathbb{E}\left[ \bar{V}\left(
x^{\prime},\varepsilon^{\prime}\right) |x,\tilde{y}\right] \right\} ,
\label{RandomAlternative}
\end{equation}%
%
%
%
where the value function $\bar{V}$ is
recursively defined via Bellman's equation as\footnote{
See, eg., Bertsekas (1987, chap. 5) for an introduction and derivation of this equation.
} 
\begin{equation*}
\bar{V}\left( x,\varepsilon \right) =\max_{\tilde{y}\in \mathcal{Y}}\left\{ 
\bar{u}_{\tilde{y}} \left(x\right) +\varepsilon _{\tilde{y}}+\beta \mathbb{E}%
\left[ \bar{V}\left( x^{\prime},\varepsilon^{\prime}\right) |x,\tilde{y}%
\right] \right\} .
\end{equation*}
}

{ $V(x)$, the ex-ante value function, is defined as:\footnote{
There is a difference between the definition of $V(x)$ and the last terms in Equation (1) above.   Here, we are considering the expectation of the value function $\bar{V}(x,\varepsilon)$ taken over the distribution of $\varepsilon|x$ (ie. holding the first argument fixed).   In the last term of Eq. (1), however, we are considering the expectation over the {\em joint} distribution of $(x',\varepsilon')|x$ (ie. holding neither argument fixed).
} 
\begin{equation*}
V\left( x\right) =\mathbb{E}\left[ \bar{V}\left( x,\varepsilon \right) |x%
\right] .
\end{equation*}
}

{ The expectation above is conditional on the current state 
$x$. In the literature, $V(x)$ is called the ex-ante (or integrated) value
function, because it measures the continuation value of the dynamic
optimization problem before the agent observes his shocks $\varepsilon $, so
that the optimal action is still stochastic from the agent's point of view. }

{ Next we define the \emph{choice-specific value functions} as
consisting of two terms: the per-period utility flow and the discounted
continuation payoff: 
\begin{equation*}
w_{y}(x) \equiv \bar{u}_{y}(x) + \beta \mathbb{E}\left[ V(x^{\prime}) |x,y)%
\right].
\end{equation*}
}
In this paper, the utility flows $\left\{u_y(x); \forall y\in\mathcal{Y}, \forall x\in\mathcal{X}\right\}$, and subsequently also the choice-specific value functions $\left\{w_y(x), \forall y,x \right\}$, will be treated as unknown parameters; and we will study the identification and estimation of these parameters.   For this reason, in the initial part of the paper, we will suppress the explicit dependence of $w_y$ on $x$ for convenience.

{ Given these preliminaries, we derive the duality which is
central to this paper. }

\subsection{\protect\normalsize The social surplus function and its convex
conjugate}

{ 
%
We start by introducing the expected indirect utility of a decision maker
facing the $|\mathbb{\mathcal{Y}}|$-dimensional vector of choice-specific
values $w\equiv \left\{w_y, y\in\mathcal{Y}\right\}'$: 
\begin{equation}
\mathcal{G}\left( w;x\right) =\mathbb{E}\left[ \max_{y\in \mathcal{Y}}\left(
w_{y}+\varepsilon _{y}\right) |x\right]  \label{swf}
\end{equation}%
where the expectation is assumed to be finite and is taken over the
distribution of the utility shocks, $Q(\cdot; x)$.    
This function $\mathcal{G}%
(\cdot ;x):\mathbb{R}^{|\mathcal{Y}|}\rightarrow \mathbb{R}$, is called the
\textquotedblleft social surplus function\textquotedblright\ in McFadden's
(1978) random utility framework, and can be interpreted as the expected
welfare of a representative agent in the dynamic discrete-choice problem. 
}

{ For convenience in what follows, we introduce the notation $%
Y(w,\varepsilon )$ to denote an agent's optimal choice given the vector of
choice-specific value functions $w$ and the vector of utility shocks $%
\varepsilon $; that is, $Y(w,\varepsilon )=\text{argmax}_{y\in \mathcal{Y}%
}(w_{y}+\varepsilon _{y})$.\footnote{{ We use $w$ and $%
\varepsilon$ (and also $p$ below) to denote vectors, while $w_{y}$ and $%
\varepsilon _{y}$ (and $p_y$) denote the $y$-th component of these vectors.}}
This notation makes explicit the randomness in the optimal alternative
(arising from the utility shocks $\varepsilon$). We get 
\begin{equation}
\mathcal{G}\left( w;x\right) =\mathbb{E}\left[ w_{Y(w,\varepsilon
)}+\varepsilon _{Y(w,\varepsilon )}|x\right] =\sum_{y\in \mathcal{Y}}%
\underbrace{{Pr}(Y(w,\varepsilon )=y|x)}_{\equiv p_y(x)}\left( w_{y}+%
\mathbb{E}[\varepsilon _{y}|Y(w,\varepsilon )=y,x]\right)  \label{ystar}
\end{equation}%
which shows an alternative expression for the social surplus function as a
weighted average, where the weights are the components of the vector of 
\emph{conditional choice probabilities} $p(x)$. For the remainder of this
section, we suppress the dependence of all quantities on $x$ for
convenience. In later sections, we will reintroduce this dependence when it
is necessary. }

{ In the case when the social surplus function $\mathcal{G}(w)$
is differentiable (which holds for most discrete-choice model specifications
considered in the literature\footnote{{ This includes logit,
nested logit, multinomial probit, etc. in which the distribution of the
utility shocks is absolutely continuous and $w$ is bounded, cf. Lemma 1 in
Shi, Shum and Wong (2014).}}), we obtain a well-known fact that the vector
of choice probabilities $p$ compatible with rational choice coincides with
the gradient of $\mathcal{G}$ at $w$: }

\begin{proposition}[The Williams-Daly-Zachary (WDZ) Theorem]
{ \label{prop:WDZ} 
\begin{equation*}
p=\nabla \mathcal{G}(w).
\end{equation*}
}
\end{proposition}

{ 
This result, which is analogous to Roy's Identity in discrete choice models,
is expounded in McFadden (1978) and Rust (1994; Thm. 3.1)). It characterizes
the vector of choice probabilities corresponding to optimal behavior in a
discrete choice model as the gradient of the social surplus function. For
completeness, we include a proof in the Appendix. The WDZ theorem provides a
mapping from the choice-specific value functions (which are unobserved by
researchers) to the observed choice probabilities $p$. }

{ However, the identification problem is the reverse
problem, namely to determine the set of $w$ which would lead to a given
vector of choice probabilities. This problem is exactly solved by convex
duality and the introduction of the convex conjugate of $\mathcal{G}$, which
we denote as $\mathcal{G}^{\ast }$:\footnote{{ Details of convex
conjugates are expounded in the Appendix. Convex conjugates are also
encountered in classic producer and consumer theory. For instance, when $f$ is the convex cost function of
the firm (decreasing returns to scale in production), then the convex
conjugate of the cost function, $f^{*}$, is in fact the firm's optimal
profit function.}} }

\begin{definition}[Convex Conjugate]
{ \label{def:convConj}We define $\mathcal{G}^{\ast }$, the
Legendre-Fenchel conjugate function of $\mathcal{G}$ (a convex function), by 
\begin{equation}
\mathcal{G}^{\ast }\left( p\right) =\sup_{w\in \mathbb{R}^{\mathcal{Y}%
}}\left\{ \sum_{y\in \mathcal{Y}}p_{y}w_{y}-\mathcal{G}\left( w\right)
\right\} .  \label{gstar}
\end{equation}
}
\end{definition}

{ Equation (\ref{gstar}) above has the property that if 
$p$ is not a probability, that is if either conditions $p_{y}\geq 0$ or $%
\sum_{y\in \mathcal{Y}}p_{y}=1$ do not hold, then $\mathcal{G}^{\ast }\left(
p\right) =+\infty $. Because the choice-specific value functions $w$ and the
choice probabilities $p$ are, respectively, the arguments of the functions $%
\mathcal{G}$ and its convex conjugate function $\mathcal{G}^{\ast }$, we say
that $w$ and $p$ are related in the sense of \emph{conjugate duality}. The
theorem below states an implication of this duality, and provides an
\textquotedblleft inverse\textquotedblright\ correspondence from the
observed choice probabilities back to the unobserved $w$, which is a
necessary step for identification and estimation. 
}

\begin{theorem}
{ \label{thm:subdiff}The following pair of equivalent statements
capture the empirical content of the DDC model:
\newline
(i) $p$ is in the subdifferential of $\mathcal{G}$ at $w$ 
\begin{equation}
p\in \partial \mathcal{G}\left( w\right) ,  \label{choiceprob}
\end{equation}%
(ii) $w$ is in the subdifferential of $\mathcal{G}^{\ast }$ at $p$ 
\begin{equation}
w\in \partial \mathcal{G}^{\ast }\left( p\right) .  \label{keyw}
\end{equation}%
}
\end{theorem}

{ The definition and properties of the subdifferential of a
convex function are provided in Appendix A.\footnote{{ $\mathcal{G%
}$ is differentiable at $w$ if and only if $\partial \mathcal{G}(w)$ is
single-valued. In that case, part (i) of Th. \ref{thm:subdiff} reduces to $%
p=\nabla \mathcal{G}(w)$, which is the WDZ theorem. If, in addition, $\nabla 
\mathcal{G}$ is one-to-one, then we immediately get $w=\left( \nabla 
\mathcal{G}\right) ^{-1}(p)$, or $\nabla \mathcal{G}^{\ast }(p)=\left(
\nabla \mathcal{G}\right) ^{-1}(p)$, which is the case of the classical
Legendre transform. However, as we show below, $\nabla \mathcal{G}(w)$ is
not typically one-to-one in discrete choice models, so that the statement in
part (ii) of Th. \ref{thm:subdiff} is more suitable.}}  Part (i) is, of
course, connected to the WDZ theorem above; indeed, it is the WDZ theorem
when $\mathcal{G}(w)$ is differentiable at $w$. Hence, it encapsulates an
optimality requirement that the vector of observed choice probabilities $p$
be derived from optimal discrete-choice decision making for some unknown
vector $w$ of choice-specific value functions. }

{ Part (ii) of this proposition, which describes the
\textquotedblleft inverse\textquotedblright\ mapping from conditional choice
probabilities to choice-specific value functions, does not appear to have
been exploited in the literature on dynamic discrete choice. It relates to
Galichon and Salani\'{e} (2012) who use convex analysis to estimate matching
games with transferable utilities. It specifically states that the vector of
choice-specific value functions can be identified from the corresponding
vector of observed choice probabilities $p$ as the subgradient of the convex
conjugate function $\mathcal{G}^{\ast }(p)$. 
Eq. (\ref{keyw}) is also constructive, and suggests a procedure for
computing the choice-specific value functions corresponding to observed
choice probabilities. We will fully elaborate this procedure in subsequent
sections\footnote{{ Clearly, Theorem~\ref{thm:subdiff} also
applies to static random utility discrete-choice models, with the $w(x)$
being interpreted as the utility indices for each of the choices. As such,
Eq. (\ref{keyw}) relates to results regarding the invertibility of the
mapping from utilities to choice probabilities in static discrete choice
models (e.g. Berry (1994); Haile, Hortacsu, and Kosenok (2008); Berry,
Gandhi, and Haile (2013)). Similar results have also arisen in the
literature on stochastic learning in games (Hofbauer and Sandholm (2002);
Cominetti, Melo and Sorin (2010)).}}. }

{ Appendix A contains additional derivations related to the
subgradient of a convex function. Specifically, it is known (Eq. (\ref%
{FenchelEq})) that $\mathcal{G}(w)+\mathcal{G}^{\ast }(p)=\sum_{y\in 
\mathcal{Y}}p_{y}w_{y}$ if and only if $p\in \partial \mathcal{G}(w)$.
Combining this with Eq. (\ref{ystar}), we obtain an alternative expression
for the convex conjugate function $\mathcal{G}^{\ast } $: 
\begin{equation}
\mathcal{G}^{\ast }(p)=-\sum_{y}p_{y}\mathbb{E}[\varepsilon
_{y}|Y(w,\varepsilon )=y],  \label{expression-gstar}
\end{equation}%
corresponding to the weighted expectations of the utility shocks $%
\varepsilon _{y}$ conditional on choosing the option $y$. It is also known
that the subdifferential $\partial \mathcal{G}^*(p)$ corresponds to the
set of maximizers in the program (\ref{gstar}) which define the conjugate
function $\mathcal{G}^*(p)$; that is, 
\begin{equation}  \label{variation}
w\in \partial\mathcal{G}^*(p)\quad\Leftrightarrow\quad w\in \text{argmax}_{w\in%
\mathbb{R}^\mathcal{Y}} \left\{ \sum_{y\in \mathcal{Y}}p_{y}w_{y}-\mathcal{G}%
\left( w\right) \right\}.
\end{equation}
Later, we will exploit this variational representation of the
subdifferential $\mathcal{G}^*(p)$ for computational purposes; cf. Section 4
below. }

\begin{example}[Logit]
{ Before proceeding, we discuss the logit model, for which the
functions and relations above reduce to familiar expressions. When the
distribution $Q$ of $\varepsilon $ obeys an extreme value type I
distribution, it follows from Extreme Value theory that $\mathcal{G}$ and $%
\mathcal{G}^{\ast }$ can be obtained in closed form\footnote{{ %
Relatedly, Arcidiacono and Miller (2011, pp. 1839-1841) discuss
computational and analytical solutions for the $\mathcal{G}^{*}$ function in
the generalized extreme value setting.}}: $\mathcal{G}\left( w\right) =\log
(\sum_{y\in \mathcal{Y}}\exp (w_{y}))+\gamma $, while $\mathcal{G}^{\ast
}\left( p\right) =\sum_{y\in \mathcal{Y}}p_{y}\log p_{y}-\gamma $ if $p$
belongs in the interior of the simplex, $\mathcal{G}^{\ast }\left( p\right)
=+\infty $ otherwise ($\gamma \approx 0.57$ is Euler's constant).
Hence in this case, $\mathcal{G}^{\ast }$ is the entropy of distribution $p$%
(see Anderson, de Palma, Thisse (1988) and references therein). 

The subdifferential of $\mathcal{G}^{\ast }$ is characterized as follows: $w\in
\partial \mathcal{G}^{\ast }\left( p\right) $ if and only if $w_{y}=\log
p_{y}-K$, for some $K\in \mathbb{R}$. In this logit case the convex
conjugate function $\mathcal{G}^{\ast }$ is the entropy of distribution $p$,
which explains why it can be called a \emph{generalized entropy} function
even in non-logit contexts.\hfill $\blacksquare $ }
\end{example}

{ 
}

\subsection{\protect\normalsize Identification}

{ \label{par:ident} }

{ It follows from Theorem~\ref{thm:subdiff} that the
identification of systematic utilities boils down to the problem of
computing the subgradient of a generalized entropy function. However, from
examining the social surplus function $\mathcal{G}$, we see that if $w\in
\partial \mathcal{G}^{\ast }\left( p\right) $, then it is also true that $%
w-K\in \partial \mathcal{G}^{\ast }\left( p\right) $, where $K\in \mathbb{R}%
^{|\mathcal{Y}|}$ is a vector taking values of $K$ across all $\mathcal{Y}$
components. 
Indeed, the choice probabilities are only affected by the differences in the
levels offered by the various alternatives. In what follows, we shall tackle
this indeterminacy problem by isolating a particular $w^{0}$ among those
satisfying $w\in \partial \mathcal{G}^{\ast }\left( p\right) $, where we
choose 
\begin{equation}
\mathcal{G}\left( w^{0}\right) =0.  \label{norma}
\end{equation}
}

{ We will impose the following assumption on the heterogeneity. }

\begin{assumption}[Full Support]
{ \label{ass:fullSupport}Assume the distribution $Q$ of the
vector of utility shocks $\varepsilon$ is such that the distribution of the
vector $\left( \varepsilon _{y}-\varepsilon _{1}\right) _{y\neq 1}$ has full
support. }
\end{assumption}

{ Under this assumption, Theorem~\ref{thm:w0} below shows that
Eq. (\ref{norma}) defines $w^{0}$ uniquely. Theorem~\ref{thm:norma} will
then show that the knowledge of $w^{0}$ allows for easy recovery of all
vectors $w$ satisfying $p\in \partial \mathcal{G}\left( w\right) $. }

\begin{theorem}
{ \label{thm:w0}Under Assumption~\ref{ass:fullSupport}, let $p$
be in the interior of the simplex $\Delta^{|\mathcal{Y}|}$, (i.e. $p_{y}>0$
for each $y$ and $\sum_{y}p_{y}=1$). Then there exists a unique $w^{0}\in
\partial \mathcal{G}^{\ast }\left( p\right) $ such that $\mathcal{G}\left(
w^{0}\right) =0$. }

{ 
}
\end{theorem}

{ The proof of this theorem is in the Appendix. Moreover, even
when Assumption~\ref{ass:fullSupport} is not satisfied, $w^{0}$ will still
be set-identified; Theorem~\ref{thm:LPIdentifiedSet} below describes the
identified set of $w^{0}$ corresponding to a given vector of choice
probabilities $p$. }

{ 
}

{ Our next result is our main tool for identification; it shows
that our choice of $w^{0}(x)$, as defined in Eq. (\ref{norma}) is without
loss of generality; it is not an additional model restriction, but merely a
convenient way of \emph{representing} all $w(x)$ in $\partial \mathcal{G}%
^{\ast }\left( p\left( x\right) \right) $ with respect to a natural and
convenient reference point.\footnote{{ This indeterminacy issue
has been resolved in the existing literature on dynamic discrete choice
models (eg. Hotz and Miller (1993), Rust (1994), Magnac and Thesmar (2002)
by focusing on the \emph{differences} between choice-specific value
functions, which is equivalent to setting $w_{y_{0}}(x) $, the
choice-specific value function for a benchmark choice $y_{0}$, equal to
zero. Compared to this, our choice of $w^{0}(x)$ satisfying $\mathcal{G}%
(w^{0}(x))=0$ is more convenient in our context, as it leads to a simple
expression for the constant $K$ (see Section \ref{empcontent}).}} }

\begin{theorem}
{ \label{thm:norma} Maintain Assumption~\ref{ass:fullSupport},
and let $K$ denote any scalar $K\in \mathbb{R}$. The set of conditions%
\begin{equation*}
w\in \partial \mathcal{G}^{\ast }\left( p\right) \text{ and }\mathcal{G}%
\left( w\right) =K
\end{equation*}%
is equivalent to 
\begin{equation*}
w_{y}=w_{y}^{0}+K,~\forall y\in \mathcal{Y}.
\end{equation*}
}
\end{theorem}

{ This theorem shows that any vector within the set $\partial 
\mathcal{G}^{\ast }\left( p\right) $ can be characterized as the sum of the
(uniquely-determined, by Theorem~\ref{thm:norma}) vector $w^{0}$ and a
constant $K\in \mathbb{R}$. As we will see below, this is our invertibility\
result for dynamic discrete choice problems, as it will imply unique
identification of the vector of choice-specific value functions
corresponding to any observed vector of conditional choice probabilities.%
\footnote{{ See Berry (1994), Chiappori and Komunjer (2010),
Berry, Gandhi, and Haile (2012), among others, for conditions ensuring the
invertibility or \textquotedblleft univalence\textquotedblright\ of demand
systems stemming from multinomial choice models, under settings more general
than the random utility framework considered here.}} }

\subsection{\protect\normalsize Empirical Content of Dynamic Discrete Choice
Model}

{ \label{empcontent} }

{ To summarize the empirical content of the model, we recall the
fact that the ex-ante value function $V$ solves the following equation%
\begin{equation*}
V\left( x\right) =\sum_{y\in \mathcal{Y}}p_{y}\left( x\right) \left( \bar{u}%
_{y}\left( x\right) +\mathbb{E}[\varepsilon _{y}|Y(w,\varepsilon
)=y,x]+\beta \sum_{x^{\prime }}p\left( x^{\prime }|x,y\right) V\left(
x^{\prime }\right) \right)
\end{equation*}%
(derived in Pesendorfer and Schmidt-Dengler (2008), among others), where we
write $p(x^{\prime }|x,y)={Pr}(x_{t+1}=x^{\prime }|x_{t}=x,y_{t}=y)$. Noting
that the choice-specific value function is just 
\begin{equation}  \label{csvf}
w_{y}(x)=\bar{u}_{y}\left( x\right) +\beta \sum_{x^{\prime }}p\left(
x^{\prime }|x,y\right) V\left( x^{\prime }\right) ,
\end{equation}%
and, comparing with Eq. (\ref{ystar}), 
\begin{equation*}
V\left( x\right) =\mathcal{G}\left( w(x);x\right) \text{ and }p\left(
x\right) \in \partial \mathcal{G}\left( w(x);x\right) .
\end{equation*}
}

{ Hence, by Theorem~\ref{thm:norma}, the true $w\left( x\right) $
will differ from $w^{0}(x)$ by a constant term $V(x)$: 
\begin{equation*}
w(x)=w^{0}\left( x\right) +V\left( x\right)
\end{equation*}%
where $w^{0}\left( x\right) $ is defined in Theorem~\ref{thm:w0}. This
result is also convenient for identification purposes, as it separates
identification of $w$ into two subproblems, the determination of $w^{0}$ and
the determination of $V$. Once $w^0$ and $V$ are known, the utility flows
are determined from Eq. (\ref{csvf}). This motivates our two-step estimation
procedure, which we describe next. }

{ 
}

\section{\protect\normalsize Estimation using the Mass Transport Approach (MTA)}

{ Based upon the derivations in the previous section, we present
a two-step estimation procedure. In the first step, we use the results from
Theorem \ref{thm:norma} to recover the vector of choice-specific value
functions $w^{0}(x)$ corresponding to each observed vector of choice
probabilities $p(x) $. In the second step, we recover the utility flow
functions $\bar{u}_{y}(x)$ given the $w^{0}(x)$ obtained from the first
step. }

\subsection{{\protect\normalsize First step\label{sec:firstStep}}}

{ In the first step, the goal is to recover the vector of
choice-specific value functions $w^{0}(x)\in \partial \mathcal{G}^{\ast
}(p(x))$ corresponding to the vector of observed choice probabilities $p(x)$
for each value of $x$. In doing this, we use Theorem 1 above and Proposition 2 below, which show how $w^{0}(x)$ belongs to the subdifferential of the
conjugate function $\mathcal{G}^{\ast }(p\left( x\right) )$. %
We delay discussing these details until Section 4. There, we will show how
this problem of obtaining $w^0(x)$ can be reformulated in terms of a class
of mathematical programming problems, the Monge-Kantorovich \emph{mass
transport} problems, which leads to convenient computational procedures.
Since this is the central component of our estimation procedure, we have
named it the \emph{mass transport approach} (MTA). }

\subsection{{\protect\normalsize Second step\label{par:secondstep}}}

{ From the first step, we obtained $w^{0}(x)$ such that $%
w(x)=w^{0}(x)+V(x)$. Now in the second step, we use the recursive structure
of the dynamic model, along with fixing one of the utility flows, to \emph{%
jointly} pin down the values of ${w}(x)$ and $V(x)$. Finally, once ${w}(x)$
and $V(x)$ are known, the utility flows can be obtained from $\bar{u}%
_{y}\left( x\right) =w_{y}(x)-\beta \mathbb{E}\left[ V(x^{\prime })|x,y%
\right] $. }

{ 
}

{ In order to nonparametrically identify $\bar{u}%
_{y}\left(x\right) $, we need to fix some values of the utility flows.
Following Bajari, Chernozhukov, Hong, and Nekipelov (2009), we fix the
utility flow corresponding to a benchmark choice $y_0$ to be constant at
zero:\footnote{{ In a static discrete-choice setting (i.e. $%
\beta=0$), this assumption would be a normalization, and without loss of
generality. In a dynamic discrete-choice setting, however, this entails some
loss of generality because different values for the utility flows imply
different values for the choice-specific value functions, which leads to
differences in the optimal choice behavior. Norets and Tang (2013) discuss
this issue in greater detail.}} }

\begin{assumption}[Fix utility flow for benchmark choice]
{ \label{ass:u0} $\forall x,\quad \bar{u}_{y_{0}}\left( x\right)
=0.$ }
\end{assumption}

{ With this assumption, we get 
\begin{equation}
0=w_{y_{0}}^{0}(x)+V\left( x\right) -\beta \mathbb{E}\left[ V\left(x^{\prime
}\right) |x,y=y_{0}\right] .  \label{implicitV}
\end{equation}
}

{ Let $W$ be the column vector whose general term is $\left(
w_{y_{0}}^{0}(x)\right) _{x\in \mathcal{X}}$, let $V$ be the column vector
whose general term is $\left( V\left( x\right) \right) _{x\in \mathcal{X}}$,
and let $\Pi^{0}$ be the $|\mathcal{X}|\times |\mathcal{X}|$ matrix whose
general term $\Pi^{0}_{ij}$ is ${Pr}\left( x_{t+1}=j|x_{t}=i,y=y_{0}\right) $%
. Equation (\ref{implicitV}), rewritten in matrix notation, is%
\begin{equation*}
W=\beta \Pi^{0} V-V
\end{equation*}%
and for $\beta <1$, matrix $I-\beta \Pi^0 $ is a diagonally dominant matrix.
Hence, it is invertible and Equation (\ref{implicitV}) becomes%
\begin{equation}
V=(\beta \Pi^{0} -I)^{-1}W.  \label{EqV}
\end{equation}
}

{ The right hand side of this equation is uniquely estimated from
the data. After obtaining $V(x)$, $\bar{u}_{y}(x)$ can be nonparametrically
identified by 
\begin{equation}
\bar{u}_{y}(x)=w_{y}^{0}(x)+V\left( x\right) -\beta \mathbb{E}[V(x^{\prime
})|x,y],  \label{EqU}
\end{equation}%
where $w^{0}\left( x\right) $ is as in Theorem~\ref{thm:norma}, and $V$ is
given by (\ref{EqV}). }

{ As a sanity check, one recovers $\bar{u}_{y_{0}}(.)=W+V-\beta
\Pi^{0} V=0$. Also, when $\beta \rightarrow 0$, one recovers $\bar{u}%
_{y}(x)=w_{y}^{0}(x)-w_{y_{0}}^{0}(x)$ which is the case in standard static
discrete choice. Moreover, since our approach to identifying the utility flows is nonparametric, our MTA approach does not leverage any known restrictions on the flow utility (including parametric or shape restrictions) in identifying or estimating the flow utilities.}\footnote{To ensure that the inverted $w$ satisfies certain shape restrictions, the linkage between $w$ and the CCP will no longer be stipulated by the subdifferential of the convex conjugate function. It is possible that there exists a modification of the convex conjugate function that is {\em equivalent} to imposing certain shape restrictions on utilities. This is an interesting avenue for future research.}


{ Eqs. (\ref{EqV}) and (\ref{EqU}) above, showing how the
per-period utility flows can be recovered from the choice-specific value
functions via a system of linear equations, echoes similar derivations in
the existing literature (e.g. Aguirregabiria and Mira (2007), Pesendorfer
and Schmidt-Dengler (2008), Arcidiacono and Miller (2011, 2013)). Hence, the
innovative aspect of our MTA estimator lies not in the second step, but
rather in the first step. 
In the next section, we delve into computational aspects of this first step. 
}

{ 
}

{ Existing procedures for estimating DDC models typically rely on
a small class of distributions for the utility shocks -- primarily those in
the extreme-value family, as in Example 1 above -- because these
distributions yield analytical (or near-analytical) formulas for the choice
probabilities and $\left\{ \mathbb{E}[\varepsilon _{y}|Y(w,\varepsilon
)=y,x]\right\} _{y}$, the vector of conditional expectation of the utility
shocks for the optimal choices, which is required in order to recover the
utility flows\footnote{{ Related papers include Hotz and Miller
(1993), Hotz, Miller, Sanders, Smith (1994), Aguirregabiria and Mira (2007),
Pesendorfer and Schmidt-Dengler (2008), Arcidiacono and Miller (2011).
Norets and Tang (2013) propose another estimation approach for binary
dynamic choice models in which the choice probability function is not
required to be known.}}. Our approach, however, which is based on computing
the $\mathcal{G}^{\ast }$ function, easily accommodates different choices
for $Q_{\varepsilon}$, the (joint) distribution of the utility shocks
conditional on $X$. Therefore, our findings expand the set of dynamic
discrete-choice models suitable for applied work far beyond those with
extreme-value distributed utility shocks.\footnote{{ This remark
is also relevant for static discrete choice models. In fact, the
random-coefficients multinomial demand model of Berry, Levinsohn, and Pakes
(1995) does not have a closed-form expression for the choice probabilities,
thus necessitating a simulation-based inversion procedure. In ongoing work
(Chiong, Galichon, Shum (2013)), we are exploring the estimation of
random-coefficients discrete-choice demand models using our approach.}} }

\section{{\protect\normalsize Computational details for the MTA estimator 
\label{sec:computation}}}

{ In Section 4.1, we show that the problem of
identification in DDC models can be formulated as a mass transport problem, and also how this may be implemented in practice. In
showing how to compute $\mathcal{G}^{\ast }$, we exploit the connection,
alluded to above, between this function and the assignment game, a model of
two-sided matching with transferable utility which has been used to model
marriage and housing markets (such as Shapley and Shubik (1971) and Becker
(1973)). 
}

\subsection{\protect\normalsize Mass Transport formulation}

{ Much of our computational strategy will be based on the
following proposition, which was derived in Galichon and Salani\'{e} (2012,
Proposition 2). It characterizes the $\mathcal{G}^{\ast }$ function as an
optimum of a well-studied mathematical program: the \textquotedblleft mass
transport,\textquotedblright problem, see Villani (2003). }

\begin{proposition}[Galichon and Salani\'{e}]
{ \label{prop:GalichonSalanie} Given Assumption (\ref%
{ass:fullSupport}), the function $\mathcal{G}^{\ast }(p)$ is the value of
the mass transport problem in which the distribution $Q$ of vectors of
utility shocks $\varepsilon $ is matched optimally to the distribution of
actions $y$ given by the multinomial distribution $p$, when the cost
associated to a match of $(\varepsilon ,y)$ is given by%
\begin{equation*}
c\left( y,\varepsilon \right) =-\varepsilon _{y}
\end{equation*}%
where $\varepsilon _{y}$ is the utility shock from taking the $y$-th action.
That is,%
\begin{equation}
\mathcal{G}^{\ast }\left( p\right) =\sup_{\substack{ w,z  \\ \text{s.t. }%
w_{y}+z\left( \varepsilon \right) \leq c\left( y,\varepsilon \right) }}%
\left\{ \mathbb{E}_{p}\left[ w_{Y}\right] +\mathbb{E}_{Q}\left[ z\left(
\varepsilon \right) \right] \right\} ,  \label{OT1}
\end{equation}%
where the supremum is taken over the pair $\left( w,z\right) $, where $w_{y}$
is a vector of dimension $\left\vert \mathcal{Y}\right\vert $ and $z(\cdot )$
is a $Q$-measurable random variable. By Monge-Kantorovich duality, (\ref{OT1}%
) coincides with its dual%
\begin{equation}
\mathcal{G}^{\ast }\left( p\right) =\min_{\substack{ Y\sim p  \\ \varepsilon
\sim Q}}\mathbb{E}\left[ c\left( Y,\varepsilon \right) \right] ,  \label{OT2}
\end{equation}%
where the minimum is taken over the joint distribution of $\left(
Y,\varepsilon \right) $ such that the the first margin $Y$ has distribution $%
p$ and the second margin $\varepsilon $ has distribution $Q$. Moreover, $%
w\in \partial \mathcal{G}^{\ast }\left( p\right) $ if and only if there
exists $z$ such that $\left( w,z\right) $ solves (\ref{OT1}). Finally, $%
w^{0}\in \partial \mathcal{G}^{\ast }\left( p\right) $ and $\mathcal{G}%
(w^{0})=0$ if and only if there exists $z$ such that $\left( w^{0},z\right) $
solves (\ref{OT1}) and $z$ is such that $\mathbb{E}_{Q}\left[ z\left(
\varepsilon \right) \right] =0$. }
\end{proposition}

{ In Eq. (\ref{OT2}) above, the minimum is taken across all joint
distributions of $(Y,\varepsilon )$ with marginal distribution equal to,
respectively, $p$ and $Q$. It follows from the proposition that the main
problem of identification of the choice-specific value functions $w$ can be
recast as a mass transport problem (Villani (2003)), in which the set of
optimizers to Eq. (\ref{OT1}) yield vectors of choice-specific value
functions $w\in \partial \mathcal{G}^{\ast }\left( p\right) $. }

{ Moreover, the mass transport problem can be interpreted as an
optimal matching problem. Using a marriage market analogy, consider a
setting in which a matched couple consisting of a \textquotedblleft
man\textquotedblright\ (with characteristics $y\sim p$) and a
\textquotedblleft woman\textquotedblright\ (with characteristics $%
\varepsilon \sim {Q}$) obtain a joint marital surplus $-c(y,\varepsilon
)=\varepsilon _{y}$. Accordingly, Eq. (\ref{OT2}) is an optimal matching
problem in which the joint distribution of characteristics $(y,\varepsilon )$
of matched couples is chosen to maximize the aggregate marital surplus. }

{ In the case when $Q$ is a discrete distribution, the mass
transport problem in the above proposition reduces to a linear-programming
problem which coincides with the assignment game of Shapley and Shubik
(1971). This connection suggests a convenient way for efficiently computing
the $\mathcal{G}^{\ast }$ function (along with its subgradient).
Specifically, we will show how the dual problem (Eq. (\ref{OT2})) takes the
form of a linear programming problem or assignment game, for which some of
the associated Lagrange multipliers correspond to the the subgradient $%
\partial \mathcal{G}^{\ast }$, and hence the choice-specific value
functions. These computational details are the focus of Section \ref%
{sec:computation} below. We include the proof of Proposition~\ref%
{prop:GalichonSalanie} in the Appendix for completeness. }

\subsection{\protect\normalsize Linear programming computation}

{ 
}

{ Let $\hat{Q}$ be a discrete approximation to the distribution $%
Q $. Specifically, consider a $S$-point approximation to $Q$, where the
support is $\text{Supp}(\hat{Q})=\{\varepsilon ^{1},\dots ,\varepsilon
^{S}\} $. Let ${Pr}(\hat{Q}=\varepsilon ^{s})=q_{s}$. The best $S$-point
approximation is such that the support points are equally weighted, $q_{s}=%
\frac{1}{S}$, i.e. the best $\hat{Q}$ is a uniform distribution, see Kennan
(2006). Therefore, let $\hat{Q}$ be a uniform distribution whose support can
be constructed by drawing $S$ points from the distribution $Q$. Moreover, 
$\hat{Q}$ converges to $Q$ uniformly as $S\rightarrow \infty $,\footnote{{
Because $\hat{Q}$ is constructed from i.i.d. draws from $Q$, this uniform convergence follows from the Glivenko-Cantelli Theorem.
}} 
so that the approximation error from this discretization will vanish when $S$
is large. Under these assumptions, Problem (\ref{OT1})-(\ref{OT2}) has a
Linear Programming formulation as%
\begin{eqnarray}
&&\max_{\pi \geq 0}\sum_{y,s}\pi _{ys}\varepsilon _{y}^{s}
\label{q discrete} \\
&&\sum_{s=1}^{S}\pi _{ys}=p_{y},\ \forall y\in \mathcal{Y}  \label{cstry} \\
&&\sum_{y\in \mathcal{Y}}\pi _{ys}=q_{s},\ \forall s\in \left\{
1,...,S\right\} .  \label{cstre}
\end{eqnarray}
}

{ For this discretized problem, the set of $w\in \partial 
\mathcal{G}^{\ast }\left( p\right) $ is the set of vectors $w$ of
Lagrange multipliers corresponding to constraints (\ref{cstry}). To see how
we recover $w^{0}$, the specific element in $\partial \mathcal{G}^{\ast
}\left( p\right) $ as defined in Theorem 1, we begin with the dual problem 
\begin{eqnarray}
&&\min_{\lambda ,z}\sum_{y\in \mathcal{Y}}p_{y}\lambda
_{y}+\sum_{s=1}^{S}q_{s}z_{s}  \label{q_discreteDual} \\
&&s.t.~\lambda _{y}+z_{s}\geq \varepsilon _{y}^{s}  \notag
\end{eqnarray}%
Consider $\left( \lambda ,z\right) $ a solution to (\ref{q_discreteDual}).
By duality, $\lambda $ and $z$ are, respectively, vectors of Lagrange
multipliers associated to constraints (\ref{cstry}) and (\ref{cstre}).%
\footnote{{ Because the two linear programs (\ref{q discrete})
and (\ref{q_discreteDual}) are dual to each other, the Lagrange multipliers
of interest $\lambda _{y}$ can be obtained by computing either program. In
practice, for the simulations and empirical application below, we computed
the primal problem (\ref{q discrete}).}} We have $\mathcal{G}^{\ast }\left(
p\right) =\sum_{y\in \mathcal{Y}}p_{y}\lambda _{y}+\sum_{s=1}^{S}q_{s}z_{s}$%
, which implies\footnote{{ This uses Eq. (\ref{FenchelEq}) in
Appendix A, which (in our setup) states that $\mathcal{G}^{*}(p)+\mathcal{G}%
(\lambda )=p\cdot \lambda $, for all Lagrange multiplier vectors $\lambda
\in \partial \mathcal{G}^{*}(p)$.}} that $\mathcal{G}\left( \lambda \right)
=-\sum_{s=1}^{S}q_{s}z_{s}$. Also, for any two elements $\lambda ,w^{0}\in
\partial \mathcal{G}^{\ast }(p)$, we have $\sum_{y\in \mathcal{Y}%
}p_{y}\lambda _{y}-\mathcal{G}(\lambda )=\sum_{y\in \mathcal{Y}%
}p_{y}w_{y}^{0}-\mathcal{G}(w^{0})$. }

{ Hence, because $\mathcal{G}(w^{0})=0$, we get%
\begin{equation}
w_{y}^{0}=\lambda _{y}-\mathcal{G}\left( \lambda \right) =\lambda
_{y}+\sum_{s=1}^{S}q_{s}z_{s}.  \label{defwon}
\end{equation}
}

{ In Theorem 5 below, we establish the consistency of this
estimate of $w^0$. }

\subsection{\protect\normalsize Discretization of $Q$ and a second type of
indeterminacy issue}

{ \label{sec:PartId} }

{ Thus far, we have proposed a procedure for computing $\mathcal{G%
}^{\ast }$ (and the choice-specific value functions $w^{0}$) by discretizing
the otherwise continuous distribution $Q$. However, because the support of $%
\varepsilon $ is discrete, $w_{y}^{0}$ will generally not be unique.%
\footnote{{ Note that Theorem 1 requires $\varepsilon $ to have
full support.}} This is due to the non-uniqueness of the solution to the
dual of the LP problem in Eq. (\ref{q discrete}), and corresponds to Shapley
and Shubik's (1971) well-known results on the multiplicity of the core in
the finite assignment game. Applied to discrete-choice models, it implies
that when the support of the utility shocks is finite, the utilities from
the discrete-choice model will only be partially identified. In this
section, we discuss this partial identification, or indeterminacy, problem
further. }

{ Recall that%
\begin{equation}
\mathcal{G}^{\ast }\left( p\right) =\sup_{w_{y}+z\left( \varepsilon \right)
\leq c\left( y,\varepsilon \right) }\left\{ \mathbb{E}_{p}\left[ w_{Y}\right]
+\mathbb{E}_{Q}\left[ z\left( \varepsilon \right) \right] \right\}
\label{dualFormulation}
\end{equation}%
where $c\left( y,\varepsilon \right) =-\varepsilon _{y}$. In Proposition~\ref%
{prop:GalichonSalanie}, this problem was shown to be the dual formulation of
an optimal assignment problem. 
}

{ We call \emph{identified set of payoff vectors}, denoted by $%
\mathcal{I}\left( p\right) $, the set of vectors $w$ such that 
\begin{equation}
\Pr \left( w_{y}+\varepsilon _{y}\geq \max_{y^{\prime }}\{w_{y^{\prime
}}+\varepsilon _{y^{\prime }}\}\right) =p_{y}  \label{choiceProb}
\end{equation}%
and we denote by $\mathcal{I}_{0}\left( p\right) $ the \emph{normalized
identified set of payoff vectors}, that is the set of $w\in \mathcal{I}%
\left( p\right) $ such that $\mathcal{G}\left( w\right) =0$. If $Q$
were to have full support, $\mathcal{I}_{0}\left( p\right) $ would contain
only the singleton $\left\{ w^{0}\right\} $ as in Theorem~\ref{thm:norma}.
Instead, when the distribution $Q$ is discrete, the set $\mathcal{I}%
_{0}\left( p\right) $ contains a multiplicity of vectors $w$ which satisfy (%
\ref{choiceprob}). One has: }

\begin{theorem}
{ \label{thm:LPIdentifiedSet}The following holds: }

{ (i) The set $\mathcal{I}\left( p\right) $ coincides with the
set of $w$ such that there exists $z$ such that $\left( w,z\right) $ is a
solution to (\ref{dualFormulation}). Thus%
\begin{equation*}
\mathcal{I}\left( p\right) =\left\{ w:\exists z,~%
\begin{array}{c}
w_{y}+z_{\varepsilon }\leq c\left( y,\varepsilon \right) \\ 
\mathbb{E}_{p}\left[ w_{Y}\right] +\mathbb{E}_{Q}\left[ z_{\varepsilon }%
\right] =\mathcal{G}^{\ast }\left( p\right)%
\end{array}%
\right\} .
\end{equation*}
}

{ (ii) The set $\mathcal{I}_{0}\left( p\right) $ is determined by
the following set of linear inequalities%
\begin{equation*}
\mathcal{I}_{0}\left( p\right) =\left\{ w:\exists z,~%
\begin{array}{c}
w_{y}+z_{\varepsilon }\leq c\left( y,\varepsilon \right) \\ 
\mathbb{E}_{p}\left[ w_{Y}\right] =\mathcal{G}^{\ast }\left( p\right) \\ 
\mathbb{E}_{Q}\left[ z_{\varepsilon }\right] =0%
\end{array}%
\right\} .
\end{equation*}
}
\end{theorem}

{ This result allows us to easily derive bounds on the individual components of $w^0$
using the characterization of the identified set using linear inequalities.
Indeed, for each $y\in \mathcal{Y}$, we can obtain upper (resp. lower)
bounds on $w_{y}$ by maximizing (resp. minimizing) $w_{y}$ subject to the
linear inequalities characterizing $\mathcal{I}_{0}(p)$,\footnote{However, letting $\bar{w}_y$ (resp. $\underline{w}_y$) denote the upper (resp. lower) bound on $w_y$, we note that typically the vector $( w_y, y\in\mathcal{Y})'\not\in \mathcal{I}_{0}(p)$.
} which is a linear
programming problem.\footnote{{ Moreover, partial identification
in $w^0$ (due to discretization of the shock distribution $Q(\varepsilon)$
will naturally also imply partial identification in the utility flows $u^0$.
For a given identified vector $w^0$ (and also given the choice probabilities 
$p$ and transition matrix $\Pi^0$ from the data), we can recover the
corresponding $u^0$ using Eqs. (\ref{EqV})-(\ref{EqU}). \label{ident u
footnote} }} }

{ Furthermore, when the dimensionality of discretization, $S$, is
high, the core shrinks to a singleton, and the core collapses to $%
\left\{ w^{0}\right\}$. This is a consequence of our next theorem, which is a consistency result.\footnote{{Gretsky, Ostroy, and Zame (1999) also discusses this phenomenon in their paper.}} In our Monte
Carlo experiments below, we provide evidence for the magnitude of this
indeterminacy problem under different levels of discretization. }

\subsection{\protect\normalsize Consistency of MTA estimator}

{ Here we show (strong) consistency for our MTA estimator of $%
w^{0}$, the normalized choice-specific value functions.  In our proof, we
accommodate two types of error: (i) approximation error from discretizing
the distribution $Q$ of $\varepsilon $, and (ii) sampling error from our
finite-sample observations of the choice probabilities. We use $Q^{n}$ to
denote the discretized distributions of $\varepsilon$, and $p^n$ to denote
the sample estimates of the choice probabilities. The limiting vector of
choice probabilities is denoted $p^0$. For a given $(Q^n, p^n)$, let $%
w_{y}^{n}$ denote the choice-specific value functions estimated using our
MTA approach. }

\begin{theorem}
{ \label{thm:consistency}Assume: }

{ (i) The sequence of vectors $\left\{ p_{y}^{n}\right\} _{y\in 
\mathcal{Y}}$, viewed as the multinomial distribution of $y$, converges
weakly to $p^{0}$; }

{ (ii) The discretized distributions of $\varepsilon $ converge
weakly to $Q$: $Q^{n}\overset{d}{\rightarrow }Q$; }

{ (iii) The second moments of $Q^{n}$ are uniformly bounded. }

{ Then the convergence $w_{y}^{n}\rightarrow w_{y}^{0}$ for each $%
y\in \mathcal{Y}$ holds almost surely. }
\end{theorem}

{ The proof, which is in the appendix, may be of independent
interest as the main argument relies on approximation results from mass
transport theory, which we believe to be the first use of such results for
proving consistency in an econometrics context. }

\section{\protect\normalsize Monte Carlo Evidence}

{ In this section, we illustrate our estimation framework using a
dynamic model of resource extraction. To illustrate how our method can
tractably handle any general distribution of the unobservables, we use a
distribution in which shocks to different choices are correlated. We will
begin by describing the setup. }

{ At each time $t$, let $x_{t} \in \{1,2,\dots,30\}$ be the state
variable denoting the size of the resource pool. There are three choices, }

\begin{description}
\item[$y_{t}=0$] { The pool of resources is extracted fully. $%
x_{t+1}|x_{t},y_{t}=0$ follows a multinomial distribution on $\{1,2,3,4\}$
with parameter $\pi =(\pi _{1},\pi _{2},\pi _{3},\pi _{4})$. The utility
flow is $\bar{u}(y_{t}=0,x_{t})=0.5\sqrt{x_{t}}-2+\varepsilon _{0}$. }

\item[$y_{t}=1$] { The pool of resources is extracted partially. $%
x_{t+1}|x_{t},y_{t}=1$ follows a multinomial distribution on $\{\max
\{1,x_{t}-10\},\max \{2,x_{t}-9\},\max \{3,x_{t}-8\},\max \{4,x_{t}-7\}\}$
with parameter $\pi $. The utility flow is $\bar{u}(y_{t}=1,x_{t})=0.4\sqrt{%
x_{t}}-2+\varepsilon _{1}$. }

\item[$y_{t}=2$] { Agent waits for the pool to grow and does not
extract. $x_{t+1}|x_{t},y_{t}=3$ follows a multinomial distribution on $%
\{x_{t},x_{t}+1,x_{t}+2,x_{t}+3\}$ with parameter $\pi $. We normalize the
utility flow to be $\bar{u}(y_{t}=2,x_{t})=\varepsilon _{2}$. }
\end{description}

{ 
}

{ The joint distribution of the unobserved state variables is
given by $(\varepsilon _{0}-\varepsilon _{2},\varepsilon _{1}-\varepsilon
_{2})\sim N\left( 
\begin{pmatrix}
0 \\ 
0%
\end{pmatrix}%
,%
\begin{pmatrix}
0.5 & 0.5 \\ 
0.5 & 1%
\end{pmatrix}%
\right) $. Other parameters we fix and hold constant for the Monte Carlo
study are the discount rate, $\beta =0.9$ and $\pi =(0.3,0.35,0.25,0.10)$. }

\subsection{\protect\normalsize Asymptotic performance}

{ As a preliminary check of our estimation procedure, we show
that we are able to recover the utility flows using the actual conditional
choice probabilities implied by the underlying model. We discretized the
distribution of $\varepsilon$ using $S=5000$ support points. As is clear
from Figure \ref{trueccp}, the estimated utility flows (plotted as dots) as
a function of states matched the actual utility functions very well. 
}

{ 
}

{ 
\begin{figure}[!h]
{ \includegraphics[scale=0.8]{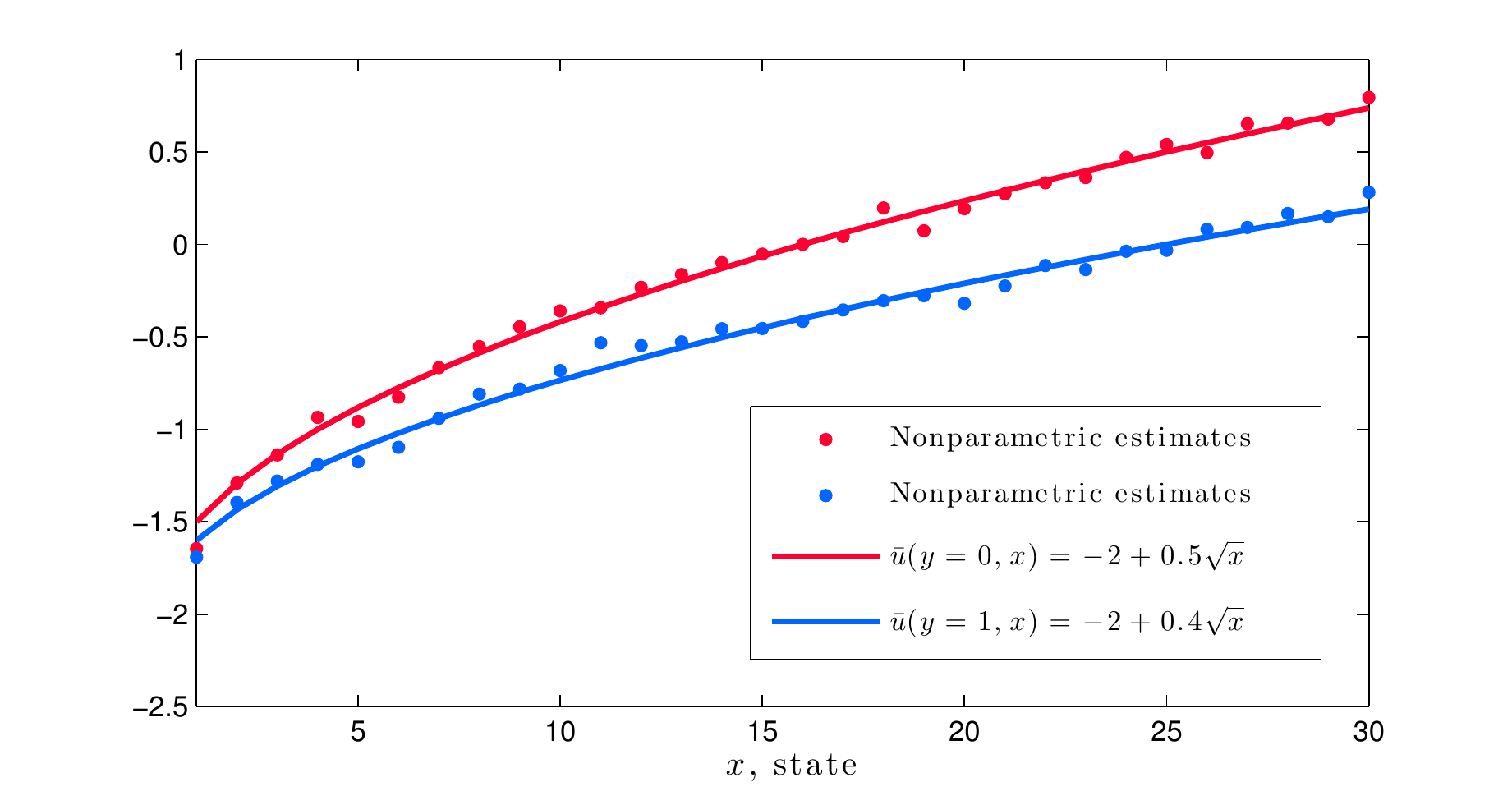}  }
\caption{Comparison between the estimated and true utility flows.}
\label{trueccp}
\end{figure}
}

\subsection{\protect\normalsize Finite sample performance}

{ To test the performance of our estimation procedure when there
is sampling error in the CCPs, we generate simulated panel data of the
following form: $\{y_{it}\;, x_{it} :i =1,2,\dots, N; \;t=1,2,\dots, T\}$
where $y_{it} \in \{0,1,2\}$ is the dynamically optimal choice at $x_{it}$
after the realization of simulated shocks. We vary the number of
cross-section observations $N$ and the number periods $T$, and for each
combination of $(N,T)$, we generate 100 independent datasets.\footnote{%
{ In each dataset, we initialized $x_{i1}$ with a random state in 
$\mathcal{X}$. When calculating RMSE and $R^{2}$, we restrict to states where the probability is in the interior of the simplex $\Delta^3$, otherwise utilities are not identified and the estimates are meaningless.}} }

{ For each replication or simulated dataset, the root-mean-square
error (RMSE) and $R^{2}$ are calculated, showing how well the estimated $%
\bar{u}_{y}(x)$ fits the true utility function for each $y$. The averages are reported in
Table \ref{mc}. }

{ 
\begin{table}[h]
{ 
\begin{tabular}{|c|c|c|c|c|}
\hline
Design & RMSE($y=0$) & RMSE($y=1$) & $R^2(y=0)$ & $R^2(y=1)$ \\ \hline
$N=100, T=100$ & 0.5586 & 0.2435 & 0.3438 & 0.7708 \\ 
$N=100, T=500$ & 0.1070 & 0.1389 & 0.7212 & 0.9119 \\ 
$N=100, T=1000$ & 0.0810 & 0.1090 & 0.8553 & 0.9501 \\ \hline
$N=200, T=100$ & 0.1244 & 0.1642 & 0.5773 & 0.8736 \\ 
$N=200, T=200$ & 0.1177 & 0.1500 & 0.7044 & 0.9040 \\ \hline
$N=500, T=100$ & 0.0871 & 0.1162 & 0.8109 & 0.9348 \\ 
$N=500, T=500$ & 0.0665 & 0.0829 & 0.8899 & 0.9678 \\ \hline
$N=1000, T=100$ & 0.0718 & 0.0928 & 0.8777 & 0.9647 \\ 
$N=1000, T=1000$ & 0.0543 & 0.0643 & 0.9322 & 0.9820 \\ \hline
\end{tabular}
}
\caption{Average fit across all replications. Standard deviations are
reported in the Appendix.}
\label{mc}
\end{table}
}

{ 
}

{ 
}

{ 
}

{ 
}

\subsection{\protect\normalsize Size of the identified set of payoffs}

As mentioned in Section~\ref{sec:PartId}, using a discrete
approximation to the distribution of the unobserved state variable
introduces a partial identification problem: the identified choice-specific
value functions might not be unique. 
Using simulations, we next show that the identified set of choice-specific
value functions (which we will simply refer to as \textquotedblleft
payoffs\textquotedblright ) shrinks to a singleton as $S$ increases, where $%
S $ is the number of support points in the discrete approximation of the continuous error distribution.
For $S$ ranging from 100 to 1000, we plot in Figure \ref{c3}, the
differences between the largest and smallest choice-specific value function
for $y=2$ across all values of $p\in \Delta ^{3}$ (using the linear
programming procedures described in Section \ref{sec:PartId}).

{ 
}

{ 
\begin{figure}[h]
\caption{The identified set of payoffs shrinks to a singleton across $%
\Delta^{3}$.}
\label{c3}{ \centering
\includegraphics[scale=0.65]{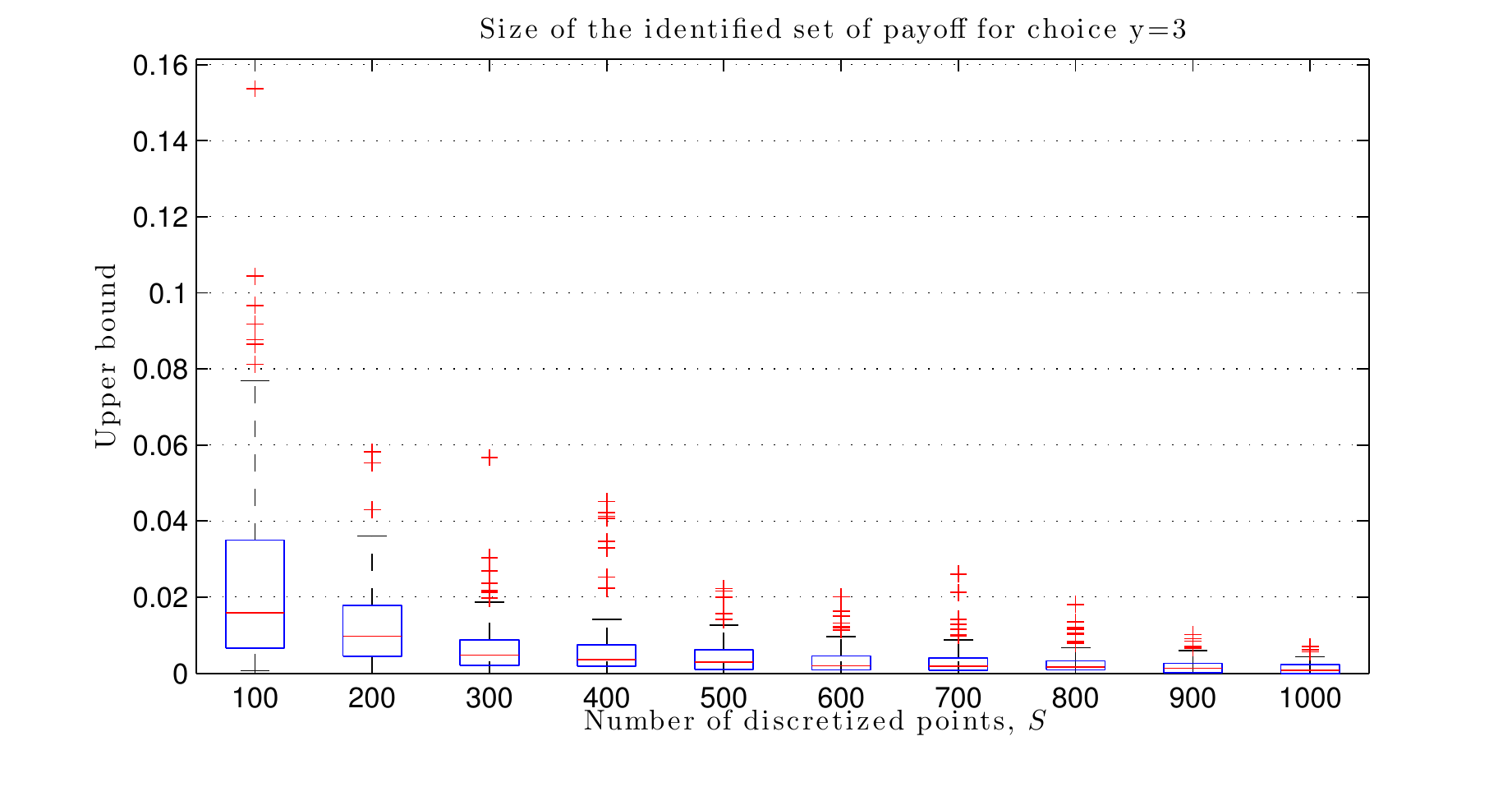} {%
\linespread{1}{\small \\ For
each value of $S$, we plot the values of the differences  $\max_{w\in
\partial\mathcal{G}^*(p)} w-\min_{w\in \partial\mathcal{G}^*(p)} w$ across
all values of $p\in\Delta^3$. In  the boxplot, the central mark is the
median, the edges of the box are  the 25th and 75th percentiles, the whiskers
extend to the most extreme data points not considered outliers, and outliers
are plotted individually.}}  }
\end{figure}
}

As is evident, even at small $S$, the identified payoffs are
very close to each other in magnitude. At $S=1000$, where computation is
near-instantaneous, for most of the values in the discretised grid of $%
\Delta^{3}$, the core is a singleton; when it is not, the difference in the
estimated payoff is less than 0.01. Similar results hold for the
choice-specific value functions for choices $y=0$ and $y=1$, which are
plotted in Figures \ref{c1} and \ref{c2} in the Appendix. To sum up, it appears that this indeterminacy issue
in the payoffs is not a worrisome problem for even very modest values of $S$.

\subsection{\protect\normalsize Comparison: MTA vs. Simulated Maximum
Likelihood}

{ One common technique used in the literature to estimate dynamic
discrete choice models with non-standard distribution of unobservables is
the Simulated Maximum Likelihood (SML). Our MTA method has a distinct
advantage over SML -- while MTA allows the utility flows $\bar{u}_{y}(x)$
for different choices $y$ and states $x$ to be nonparametric, the SML
approach typically requires parameterizing these utility flows as a function
of a low-dimensional parameter vector. This makes comparison of these two
approaches awkward. Nevertheless, here we undertake a comparison of the
nonparametric MTA vs. the parametric SML approach. First we compare
the performance of the two alternative approaches in terms of computational
time. The computations were performed on a Quad Core Intel Xeon 2.93GHz UNIX
workstation, and the results are presented in Table~\ref{smm}. 
}

From a computational point of view, the disadvantage of SML is
that the dynamic programming problem must be solved (via Bellman function
iteration) for each trial parameter vector, whereas the MTA requires solving
a large-scale linear programming problem -- but only \emph{once}. Table \ref%
{smm} shows that our MTA procedure significantly outperforms SML in terms of computational speed. This finding,
along with the results in Table \ref{mc}, show that MTA has the desirable
properties of speed and accuracy, and also allows for nonparametric
specification of the utility flows $\bar{u}_{y}(x)$.

{ 
\begin{table}[!h]
\caption{Comparison: MTA vs. Simulated Maximum Likelihood (SML)}
\label{smm}{ \centering
}
\par
{ 
\begin{tabular}{*3c}
\toprule
 \multicolumn{1}{c}{$S$ discretized points}& \multicolumn{1}{c}{SML:${}^+$
} & \multicolumn{1}{c}{MTA:${}^{++}$} \\
 \multicolumn{1}{c}{}& \multicolumn{1}{c}{Avg. seconds} & \multicolumn{1}{c}{Avg. seconds} \\
\midrule
   2000  &    19.8 & 2.6\\
   3000   &  24.5 & 4.4 \\
   4000    & 26.5 & 6.6\\
   5000    &  40.9 & 9.6\\
   6000    &  70.5 & 13.4\\
   7000    &  105.0 & 17.5 \\
   8000    &   129.4 & 21.5 \\
\bottomrule
\end{tabular}
}

${}^+$:{\footnotesize
In this column we report time it takes to estimate the parameters $\theta=(\theta_{00},\theta_{01},\theta_{10},\theta_{11})$ as a local maximum of a simulated maximum likelihood, where $\theta$ corresponds to  $\bar{u}_{y=0}(x) = \protect\theta_{00} +
\protect\theta_{01} \protect\sqrt{x}$, and $\bar{u}_{y=1}(x) = \protect\theta_{10} + \protect\theta_{11} \protect\sqrt{x}$.}\\
${}^{++}$:{\footnotesize
In this column we report the
     time it takes to nonparametrically estimate the per-period utility flow.}
\end{table}
}



Furthermore, as confirmed in our computations, the nonlinear optimization routines typically used to implement SML have trouble finding the global optimum; in contrast, the MTA estimator, by virtue of its being a linear programming problem, always finds the global optimum.  Indeed, under the logistic assumption on unobservables and linear-in-parameters utility, one advantage of the Hotz-Miller estimator for DDC models (vs. SML) is that the system of equations defining the estimator has a unique global solution; in their discussion of this, Aguirregabiria and Mira (2010, pg. 48) remark that ``extending the range of applicability of ... CCP methods to models which do not impose the CLOGIT [logistic] assumption is a topic for further research." This paper fills the gap: our MTA estimator shares the computational advantages of the CLOGIT setup, but works for non-logistic models. In this sense, the MTA estimator is a generalized CCP estimator.

\section{\protect\normalsize Empirical Application: Revisiting Harold Zurcher%
}

{ In this section, we apply our estimation procedure to the bus
engine replacement dataset first analyzed in Rust (1987). In each week $t$,
Harold Zurcher (bus depot manager), chooses $y_{t} \in \{0,1\}$ after
observing the mileage $x_{t} \in \mathcal{X}$ and the realized shocks $%
\varepsilon_{t}$. If $y_{t}=0$, then he chooses not to replace the bus
engine, and $y_{t}=1$ means that he chooses to replace the bus engine. The
states space is $\mathcal{X}=\{0,1,\dots 29\}$, that is, we divided the
mileage space into 30 states, each representing a 12,500 increment in
mileage since the last engine replacement.\footnote{{ \label%
{fn:5000} This grid is coarser compared to Rust's (1987) original analysis
of this data, in which he divided the mileage space into increments of 5,000
miles. However, because replacement of engines occurred so infrequently
(there were only 61 replacement in the entire ten-year sample period), using
such a fine grid size leads to many states that have zero probability of
choosing replacement. Our procedure -- like all other CCP-based approaches
-- fails when the vector of conditional choice probability lies on the
boundary of the simplex.}} Harold Zurcher manages a fleet of 104 identical
buses, and we observe the decisions that he made, as well as the
corresponding bus mileage at each time period $t$. The duration between $t+1 
$ and $t$ is a quarter of a year, and the dataset spans 10 years. Figures %
\ref{fig:numreplace} and \ref{fig:ccp} in the Appendix summarize the
frequencies and mileage at which replacements take place in the dataset. }

{ Firstly, we can directly estimate the probability of choosing
to replace and not to replace the engine for each state in $\mathcal{X}$. 
Also directly obtained from the data is the Markov transition probabilities
for the observed state variable $x_{t} \in X$, estimated as: }

{ 
\begin{equation*}
\hat{\text{Pr}}(x_{t+1} =j |x_{t}=i, y_{t}=0) = 
\begin{cases}
0.7405 & \mbox{if} \quad j=i \\ 
0.2595 & \mbox{if} \quad j=i+1 \\ 
0 & \mbox{otherwise}%
\end{cases}%
\end{equation*}
}

{ 
\begin{equation*}
\hat{\text{Pr}}(x_{t+1} =j |x_{t}=i, y_{t}=1) = 
\begin{cases}
0.7405 & \mbox{if} \quad j=0 \\ 
0.2595 & \mbox{if} \quad j=1 \\ 
0 & \mbox{otherwise}%
\end{cases}%
\end{equation*}
}

{ 
%
}

{ 
\begin{figure}[tbp]
{ \centering
\includegraphics[scale=0.8]{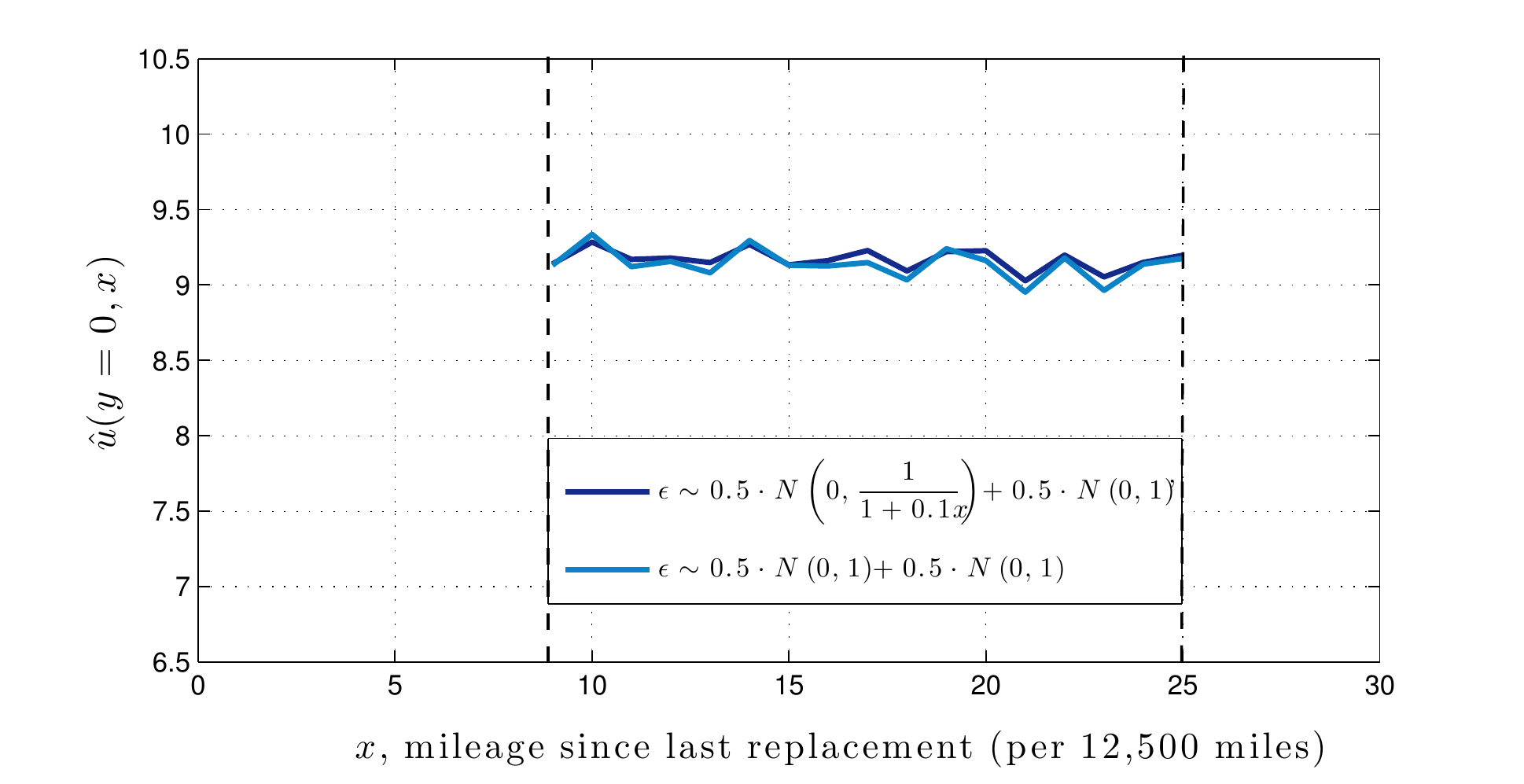}  }
\caption{Estimates of utility flows $\bar{u}_{y=0}(x)$, across values of
mileage $x$}
\label{fig:mix}
\end{figure}
}

{ For this analysis, we assumed a normal mixture distribution of
the error term, specifically, $\varepsilon_{t0} - \varepsilon_{t1} \sim 
\frac{1}{2}N(0,1)+\frac{1}{2}N(0,\frac{1}{1+0.1x})$.\footnote{{ %
In this paper, we restrict attention to the case where the researcher fully
knows the distribution of the unobservables $Q_{\vec{\varepsilon}}$, so that
there are no unknown parameters in these distributions. In principle, the
two-step procedure proposed here can be nested inside an additional ``outer
loop'' in which unknown parameters of $Q_{\vec{\varepsilon}}$ are
considered, but identification and estimation in this case must rely on
additional model restrictions in addition to those considered in this paper.
We are currently exploring such a model in the context of the simpler static
discrete choice setting (Chiong, Galichon and Shum (2014, work in progress)).%
}} We chose this mixture distribution in order to allow the utility shocks
to depend on mileage -- which accommodates, for instance, operating costs
which may be more volatile and unpredictable at different levels of mileage.
At the same time, these specifications for the utility shock distribution
showcase the flexibility of our procedure in estimating dynamic discrete
choice models for any general error distribution. For comparison, we repeat
this exercise using an error distribution that is homoskedastic, i.e., its
variance does not depend on the state variable $x_{t}$. The result appears
to be robust to using different distributions of $\varepsilon_{t0} -
\varepsilon_{t1}$. We set the discount rate $\beta=0.9$. }

{ 
}

{ To non-parametrically estimate $\bar{u}_{y=0} (x)$, we fixed $%
\bar{u}_{y=1}(x)$ to 0 for all $x \in X$. Hence, our estimates of $\bar{u}%
_{y=0}(x)$ should be interpreted as the magnitude of operating costs%
\footnote{{ Operating costs include maintenance, fuel, insurance
costs, plus Zurcher's estimate of the costs of lost ridership and goodwill
due to unexpected breakdowns.}} relative to replacement costs\footnote{%
{ To be pedantic, this also includes the operating cost at $x=0$.}%
}, with positive values implying that replacement costs exceed operating
costs. The estimated utility flows from choosing $y=0$ (don't replace)
relative to $y=1$ (replace engine) are plotted in Figure \ref{fig:mix}. We
only present estimates for mileage within the range $x\in[9,25]$, because
within this range, the CCPs are in the interior of the probability simplex
(cf. footnote \ref{fn:5000} and Figure \ref{fig:ccp} in appendix). }

{ Within this range, the estimated utility function does not vary
much with increasing mileages, i.e. it has slope that is not significantly
different from zero. The recovered utilities fall within the narrow band of
9 and 9.5, which implies that on average the replacement cost is much higher
than the maintenance cost, by a magnitude of 18 to 19 times the variance of
the utility shocks. It is somewhat surprising that our results suggest that
when the mileage goes beyond the cutoff point of 100,000 miles, Harold
Zurcher perceived the operating costs to be inelastic with respect to
accumulated mileage. It is worth noting that Rust (1987) mentioned:
``According to Zurcher, monthly maintenance costs increase very slowly as a
function of accumulated mileage." 
}

{ 
\begin{figure}[h]
\caption{Bootstrapped estimates of utility flows $\bar{u}_{y=0}(x)$}
\label{boo}{ \centering
\includegraphics[scale=0.75]{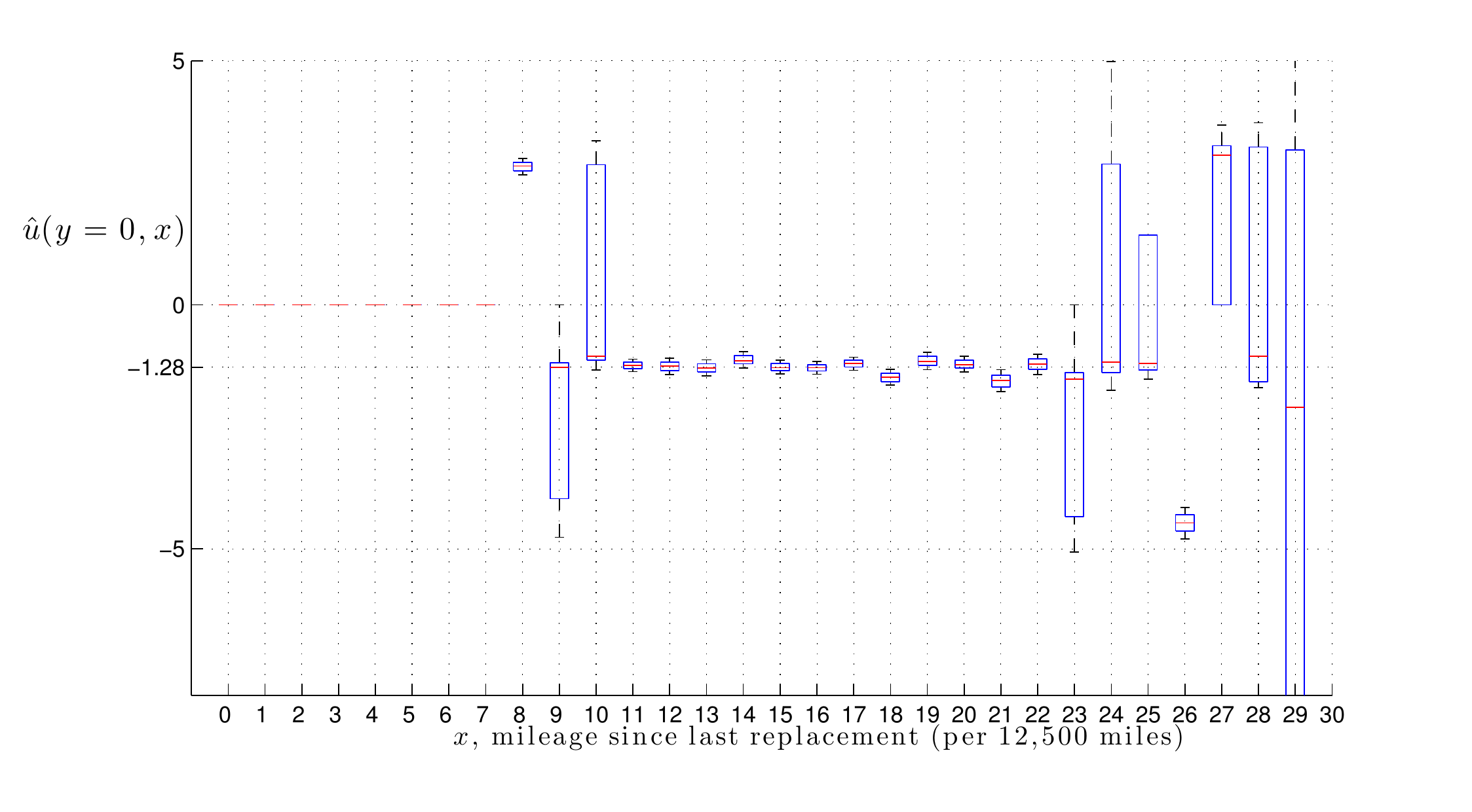} {%
\linespread{1}{\small We plot
the values of the bootstrapped resampled estimates of  $\bar{u}_{y=0}(x)$. In
each boxplot, the central mark is the median, the edges of the box are the
25th and 75th percentiles, the whiskers extend to the 5th and 95th
percentiles.}}  }
\end{figure}
}

{ To get an idea for the effect of sampling error on our
estimates, we bootstrapped our estimation procedure. For each of 100
resamples, we randomly drew 80 buses with replacement from the dataset, and
re-estimated the utility flows $\bar{u}_{y=0}(x)$ using our procedure. The
results are plotted in Figure \ref{boo}. The evidence suggests that we are
able to obtain fairly tight cost estimates for states where there is at
least one replacement, i.e. for $x\geq9$ ($x \geq 112,500$ miles), and for
states that are reached often enough; i.e. for $x\leq 22$ ($x \leq 275,000$
miles). 
}

{ 
%
}

\section{\protect\normalsize Conclusion}

{ In this paper, we have shown how results from convex analysis
can be fruitfully applied to study identification in dynamic discrete choice
models; modulo the use of these tools, a large class of dynamic discrete
choice problems with quite general utility shocks becomes no more difficult
to compute and estimate than the Logit model encountered in most empirical
applications. This has allowed us to provide a natural and holistic
framework encompassing the papers of Rust (1987), Hotz and Miller (1993),
and Magnac and Thesmar (2002). While the identification results in this
paper are comparable to other results in the literature, the approach we
take, based on the convexity of the social surplus function $\mathcal{G}$
and the resulting duality between choice probabilities and choice-specific
value functions, appears new. Far more than providing a mere reformulation,
this approach is powerful, and has significant implications in several
dimensions. }

{ First, by drawing the (surprising) connection between the
computation of the $\mathcal{G}^{\ast }$ function and the computation of
optimal matchings in the classical assignment game, 
we can apply the powerful tools developed to compute optimal matchings to
dynamic discrete-choice models.\footnote{{ While the present
paper has used standard Linear Programming algorithms such as the Simplex
algorithm, other, more powerful matching algorithms such as the Hungarian
algorithm may be efficiently put to use when the dimensionality of the
problem grows.}} Moreover, by reformulating the problem as an optimal
matching problem, all existence and uniqueness results are inherited from
the theory of optimal transport. For instance, the uniqueness of a
systematic utility rationalizing the consumer's choices follows from the
uniqueness of a potential in the Monge-Kantorovich theorem. 
}

{ We believe the present paper opens a more flexible way to deal
with discrete choice models. While identification is exact for a fixed
structure of the unobserved heterogeneity, one may wish to parameterize the
distribution of the utility shocks and do inference on that parameter. The
results and methods developed in this paper may also extend to dynamic
discrete games, with the utility shocks reinterpreted as players' private
information.\footnote{{ See, e.g. Aguirregabiria and Mira (2007)
or Pesendorfer and Schmidt-Dengler (2008)).}} However, we leave these
directions for future exploration. }

{ \appendix
}

\section{\protect\normalsize Background results}

\subsection{{\protect\normalsize Convex Analysis for Discrete-choice Models 
\label{app:convexAnalysis}}}

{ Here, we give a brief review of the main notions and results
used in the paper. We keep an informal style and do not give proofs, but we
refer to Rockafellar (1970) for an extensive treatment of the subject. }

{ Let $u\in \mathbb{R}^{|\mathcal{Y}|}$ be a vector of utility
indices. For utility shocks $\{\varepsilon _{y}\}_{y\in \mathcal{Y}}$
distributed according to a joint distribution function $Q$, we define the
social surplus function as 
\begin{equation}
\mathcal{G}(u)={\mathbb{E}}[\max_{y}{\{u_{y}+\varepsilon _{y}}\}],
\label{SSdiscreteChoice}
\end{equation}%
where $u_{y}$ is the $y$-th component of $u$. If ${\mathbb{E}}(\varepsilon
_{y}) $ exists and is finite, then the function $\mathcal{G}$ is a proper
convex function that is continuous everywhere. Moreover assuming that $Q$ is
sufficiently well-behaved (for instance, if it has a density with respect to
the Lebesgue measure), $\mathcal{G}$ is differentiable everywhere. }

{ Define the \emph{Legendre-Fenchel conjugate}, or \emph{convex
conjugate} of $\mathcal{G}$ as $\mathcal{G}^{\ast }(p)=\sup_{u\in \mathbb{R}%
^{|\mathcal{Y}|}}\{p\cdot u-\mathcal{G}(u)\}$. Clearly, $\mathcal{G}^{\ast }$
is a convex function as it is the supremum of affine functions. Note that
the inequality%
\begin{equation}
\mathcal{G}(u)+\mathcal{G}^{\ast }(p)\geq p\cdot u  \label{FenchelIneq}
\end{equation}%
holds in general. The domain of $\mathcal{G}^{\ast }$ consists of $p\in 
\mathbb{R}^{|\mathcal{Y}|}$ for which the supremum is finite. In the case
when $\mathcal{G}$ is defined by (\ref{SSdiscreteChoice}), it follows from
Norets and Takahashi (2013) that the domain of $\mathcal{G}^{\ast }$
contains the simplex $\Delta ^{|\mathcal{Y}|}$, which is the set of $p\in 
\mathbb{R}^{|\mathcal{Y}|}$ such that $p_{y}\geq 0$ and $\sum_{y\in \mathcal{%
Y}}p_{y}=1$. This means that our convex conjugate function is always
well-defined. }

{ The \emph{subgradient} $\partial \mathcal{G}\left( u\right) $
of $\mathcal{G} $ at $u$ is the set of $p\in \mathbb{R}^{|\mathcal{Y}|}$
such that 
\begin{equation*}
p\cdot u-\mathcal{G}(u)\geq p\cdot u^{\prime }-\mathcal{G}(u^{\prime })
\end{equation*}%
holds for all $u^{\prime }\in \mathbb{R}^{|\mathcal{Y}|}$. Hence $\partial 
\mathcal{G}$ is a set-valued function or correspondence. $\partial \mathcal{G%
}\left( u\right) $ is a singleton if and only if $\mathcal{G}(u)$ is
differentiable at $u$; in this case, $\partial \mathcal{G}\left( u\right)
=\nabla \mathcal{G}\left( u\right) $. }

{ One sees that $p\in \partial \mathcal{G}\left( u\right) $ if
and only if $p\cdot u-\mathcal{G}(u)=\mathcal{G}^{\ast }(p)$, that is if
equality is reached in inequality (\ref{FenchelIneq}): 
\begin{equation}
\mathcal{G}(u)+\mathcal{G}^{\ast }(p)=p\cdot u.  \label{FenchelEq}
\end{equation}
This equation is itself of interest, and is known in the literature as
``Fenchel's equality''. By symmetry in (\ref{FenchelEq}), one sees that $%
p\in \partial \mathcal{G}\left( u\right) $ if and only if $u\in \partial 
\mathcal{G}^{\ast }(p)$. In particular, when both $\mathcal{G}$ and $%
\mathcal{G}^{\ast }$ are differentiable, then $\nabla \mathcal{G}^{\ast
}=\nabla \mathcal{G}^{-1}$. }

\section{\protect\normalsize Proofs}

\begin{proof}[Proof of Proposition \protect\ref{prop:WDZ}]
{ Consider the $y$-th component, corresponding to $\frac{\partial 
\mathcal{G}(w)}{\partial w_{y}}$: 
\begin{align}
\frac{\partial \mathcal{G}(w)}{\partial w_{y}}& =\frac{\partial }{\partial
w_{y}}\int \max_{y}[w_{y}+\varepsilon _{y}]dQ \\
& =\int \frac{\partial }{\partial w_{y}}\max_{y}[w_{y}+\varepsilon _{y}]dQ \\
& =\int \mathbbm{1}(w_{y}+\varepsilon _{y}\geq w_{y^{\prime }}+\varepsilon
_{y^{\prime }}),\forall y^{\prime }\neq y)dQ=p(y).
\end{align}%
(We have suppressed the dependence on $x$ for convenience.) }
\end{proof}

\begin{proof}[Proof of Theorem \protect\ref{thm:subdiff}]
{ This follows directly from Fenchel's equality (see Rockafellar
(1970), Theorem 23.5, see also Appendix~\ref{app:convexAnalysis}), which
states that%
\begin{equation*}
p\in \partial \mathcal{G}\left( w\right)
\end{equation*}%
is equivalent to $\mathcal{G}\left( w\right) +\mathcal{G}^{\ast }\left(
p\right) =\sum_{y}p_{y}w_{y}$, which is equivalent in turn to%
\begin{equation*}
w\in \partial \mathcal{G}^{\ast }\left( p\right) .
\end{equation*}
}
\end{proof}

\begin{proof}[Proof of Theorem~\protect\ref{thm:w0}]
{ Because $\varepsilon $ has full support, the choice
probabilities $p$ will lie strictly in the interior of the simplex $\Delta
^{|\mathcal{Y}|}$. Let $\tilde{w}\in \partial \mathcal{G}^{\ast }\left(
p\right) $, and let $w_{y}=\tilde{w}_{y}-\mathcal{G}\left( \tilde{w}\right) $%
. One has $\mathcal{G}\left( w\right) =0$, and an immediate calculation
shows that $\partial \mathcal{G}\left( w\right) =p$. Let us now show that $w$
is unique. Consider $w$ and $w^{\prime }$ such that $\mathcal{G}\left(
w\right) =\mathcal{G}\left( w^{\prime }\right) =0$, and $p\in \partial 
\mathcal{G}\left( w\right) $ and $p\in \partial \mathcal{G}\left( w^{\prime
}\right) $. Assume $w\neq w^{\prime }$ to get a contradiction; then there
exist two distinct $y_{0}$ and $y_{1}$ such that $w_{y_{0}}-w_{y_{1}}\neq
w_{y_{0}}^{\prime }-w_{y_{1}}^{\prime }$; without loss of generality one may
assume%
\begin{equation*}
w_{y_{0}}-w_{y_{1}}>w_{y_{0}}^{\prime }-w_{y_{1}}^{\prime }.
\end{equation*}%
Let $S$ be the set of $\varepsilon $'s such that 
\begin{eqnarray*}
&&w_{y_{0}}-w_{y_{1}}>\varepsilon _{y_{1}}-\varepsilon
_{y_{0}}>w_{y_{0}}^{\prime }-w_{y_{1}}^{\prime } \\
&&w_{y_{0}}+\varepsilon _{y_{0}}>\max_{y\neq y_{0},y_{1}}\left\{
w_{y}+\varepsilon _{y}\right\} \\
&&w_{y_{1}}^{\prime }+\varepsilon _{y_{1}}>\max_{y\neq y_{0},y_{1}}\left\{
w_{y}^{\prime }+\varepsilon _{y}\right\}
\end{eqnarray*}%
Because $\varepsilon $ has full support, $S$ has positive probability. }

{ Let $\bar{w}=\frac{w+w^{\prime }}{2}$. Because $p\in \partial 
\mathcal{G}\left( w\right) $ and $p\in \partial \mathcal{G}\left( w^{\prime
}\right) $, $\mathcal{G}$ is linear on the segment $\left[ w,w^{\prime }%
\right] $, thus \ $\mathcal{G}\left( \bar{w}\right) =0$, thus%
\begin{eqnarray*}
0 &=&\mathbb{E}\left[ \bar{w}_{Y\left( \bar{w},\varepsilon \right)
}+\varepsilon _{Y\left( \bar{w},\varepsilon \right) }\right] =\frac{1}{2}%
\mathbb{E}\left[ w_{Y\left( \bar{w},\varepsilon \right) }+\varepsilon
_{Y\left( \bar{w},\varepsilon \right) }\right] +\frac{1}{2}\mathbb{E}\left[
w_{Y\left( \bar{w},\varepsilon \right) }^{\prime }+\varepsilon _{Y\left( 
\bar{w},\varepsilon \right) }\right] \\
&\leq &\frac{1}{2}\mathbb{E}\left[ w_{Y\left( w,\varepsilon \right)
}+\varepsilon _{Y\left( w,\varepsilon \right) }\right] +\frac{1}{2}\mathbb{E}%
\left[ w_{Y\left( w^{\prime },\varepsilon \right) }^{\prime }+\varepsilon
_{Y\left( w^{\prime },\varepsilon \right) }\right] \\
&=&\frac{1}{2}\left( \mathcal{G}\left( w\right) +\mathcal{G}\left( w^{\prime
}\right) \right) =0
\end{eqnarray*}%
}

{ Hence equality holds term by term, and 
\begin{eqnarray*}
w_{Y\left( w,\varepsilon \right) }+\varepsilon _{Y\left( w,\varepsilon
\right) } &=&w_{Y\left( \bar{w},\varepsilon \right) }+\varepsilon _{Y\left( 
\bar{w},\varepsilon \right) } \\
w_{Y\left( w^{\prime },\varepsilon \right) }^{\prime }+\varepsilon _{Y\left(
w^{\prime },\varepsilon \right) } &=&w_{Y\left( \bar{w},\varepsilon \right)
}^{\prime }+\varepsilon _{Y\left( \bar{w},\varepsilon \right) }
\end{eqnarray*}
}

{ For $\varepsilon \in S$, $Y\left( w,\varepsilon \right)
=Y\left( \bar{w},\varepsilon \right) =y_{0}$ and $Y\left( w^{\prime
},\varepsilon \right) =Y\left( \bar{w},\varepsilon \right) =y_{1}$, and we
get the desired contradiction. }

{ Hence $w=w^{\prime }$, and the uniqueness of $w$ follows. }
\end{proof}

{ 
}

{ \bigskip }

\begin{proof}[Proof of Theorem \protect\ref{thm:norma}]
{ From{\ $\mathcal{G}\left( w^{0}\right) =0$} and $\partial 
\mathcal{G}\left( w-\mathcal{G}\left( w\right) \right) =\partial \mathcal{G}%
\left( w\right) $, and by the uniqueness result in Theorem~\ref{thm:w0}, it
follows that 
\begin{equation*}
w^{0}=w-\mathcal{G}\left( w\right) .
\end{equation*}
}
\end{proof}

\begin{proof}[Proof of Proposition \protect\ref{prop:GalichonSalanie}]
{ The proof is in Galichon and Salani\'{e} (2012), but we include
it here for self-containedness. This connection between the $\mathcal{G}%
^{\ast }$ function and a matching model follows from manipulation of the
variational problem in the definition of $\mathcal{G}^{\ast }$: 
\begin{eqnarray}
\mathcal{G}^{\ast }\left( p\right) &=&\sup_{w\in \mathbb{R}^{\mathcal{Y}%
}}\left\{ \sum_{y}p_{y}w_{y}-\mathbb{E}_{Q}\left[ \max_{y\in \mathcal{Y}%
}\left( w_{y}+\varepsilon _{y}\right) \right] \right\}  \label{gstar2} \\
&=&\sup_{w\in \mathbb{R}^{\mathcal{Y}}}\left\{ \sum_{y}p_{y}w_{y}+\mathbb{E}%
_{Q}\underset{\equiv z(\varepsilon )}{\underbrace{\left[ \min_{y\in \mathcal{%
Y}}\left( -w_{y}-\varepsilon _{y}\right) \right] }}\right\} .  \notag
\end{eqnarray}%
Defining $c\left( y,\varepsilon \right) \equiv -\varepsilon _{y}$, one can
rewrite the above as 
\begin{equation}
\mathcal{G}^{\ast }\left( p\right) =\sup_{w_{y}+z\left( \varepsilon \right)
\leq c\left( y,\varepsilon \right) }\left\{ \mathbb{E}_{p}\left[ w_{Y}\right]
+\mathbb{E}_{Q}\left[ z\left( \varepsilon \right) \right] \right\} .
\label{dual}
\end{equation}%
As is well-known from the results of Monge-Kantorovich (Villani (2003), Thm.
1.3), this is the dual-problem for a mass transport problem. The
corresponding primal problem is 
\begin{equation*}
\mathcal{G}^{\ast }\left( p\right) =\min_{\substack{ Y\sim p  \\ \varepsilon
\sim \hat{Q}}}\mathbb{E}\left[ c\left( Y,\varepsilon \right) \right]
\end{equation*}%
which is equivalent to (\ref{q discrete})-(\ref{cstre}). Comparing Eqs. (\ref%
{gstar2}) and (\ref{dual}), we see that the subdifferential $\partial {%
\mathcal{G}}^{\ast }(p)$ is identified with those elements $w$ such that $%
(w,z)$, for some $z$, solves the dual problem (\ref{dual}). }
\end{proof}

\begin{proof}[Proof of Theorem \protect\ref{thm:LPIdentifiedSet}]
{ (i) follows from Proposition \ref{prop:GalichonSalanie} and the
fact that if $w_{y}+z\left( \varepsilon \right) \leq c\left( y,\varepsilon
\right) $, then $\mathbb{E}_{p}\left[ w_{Y}\right] +\mathbb{E}_{Q}\left[
z\left( \varepsilon \right) \right] =\mathcal{G}^{\ast }\left( p\right) $ if
and only if $\left( w,z\right) $ is a solution to the dual problem. }

{ (ii) follows from the fact that $-z\left( \varepsilon \right)
=\sup_{y}\left\{ w_{y}-c\left( y,\varepsilon \right) \right\}
=\sup_{y}\left\{ w_{y}+\varepsilon _{y}\right\} $, thus $\mathbb{E}_{Q}\left[
z\left( \varepsilon \right) \right] =0$ is equivalent to $\mathbb{E}_{Q}%
\left[ \sup_{y}\left\{ w_{y}+\varepsilon _{y}\right\} \right] =0$, that is $%
\mathcal{G}\left( w\right) =0$. }
\end{proof}

\begin{proof}[Proof of Theorem~\protect\ref{thm:consistency}]
{ We shall show that the vector of choice-specific value
functions derived from the MTA estimation procedure, denoted $w^{n}$,
converges to the true vector $w^{0}$. In our procedure, there are two
sources of estimation error. The first is the sampling error in the vector
of choice probabilities, denoted ${p}^{n}$. The second is the simulation
error involved in the discretization of the distribution of $\varepsilon $;
we let $Q^{n}$ denote this discretized distribution. }

{ A distinctive aspect of our proof is that it utilizes the
theory of mass transport; namely convergence results for sequences of mass
transport problems. For $y\in \mathcal{Y}$, let $\iota ^{y}$ denote the $|%
\mathcal{Y}|$-dimensional row vector with all zeros except a 1 in the $y$-th
column. This discretized mass transport problem from which we obtain $w^{n}$
is: 
\begin{equation}
\sup_{\gamma \in \mathcal{M}(Q^{n},{p}^{n})}\int_{{\mathbb{R}}^{d}\times {%
\mathbb{R}}^{d}}\left( \iota\cdot\varepsilon \right) \gamma (d\varepsilon
,d\iota )  \label{mkn}
\end{equation}%
where $\mathcal{M}(Q^{n},{p}^{n})$ denotes the set of joint (discrete)
probability measures with marginal distributions $Q^{n}$ and $p^{n}$. In the
above, $\iota $ denotes a random vector which is equal to $\iota ^{y}$ with
probability $p_{y}^{n}$, for $y\in \mathcal{Y}$. The dual problem used in
the MTA procedure is%
\begin{eqnarray}
\inf_{z,w}&&\int z\left( \varepsilon \right) dQ^{n}\left( \varepsilon
\right) +\sum_{y}w_{y}p_{y}^{n}\;:  \label{dual1} \\
s.t.~ && z(\varepsilon )\geq \iota^{y}\cdot\varepsilon -w_{y},\;\forall
y,\;\forall \varepsilon \\
&& \mathcal{G}_{n}(w_{y}^{n})=0,  \label{normass}
\end{eqnarray}%
where $\mathcal{G}_{n}(w)\equiv \mathbb{E}_{Q^{n}}(w_{y}+\epsilon _{y})$. We
let $(z^{n},w^{n})$ denote solutions to this discretized dual problem $(\ref%
{dual1})$. Recall (from the discussion in Section~\ref{par:ident}) that the
extra constraint (\ref{normass}) in the dual problem just selects among the
many dual optimizing arguments $(w^n, z^n)$ corresponding to the optimal
primal solution $\gamma^n$, and so does not affect the primal problem.%
\footnote{{ We note that, as discussed before, the discreteness
of $Q^n$ implies that $(z^{n},w^{n})$ will not be uniquely determined, as
the core of the assignment game for a finite market is not a singleton. But
this does not affect the proof, as our arguments below hold for any sequence
of selections $\left\{ z^{n},w^{n}\right\} _{n}$.}} }

{ 
Next we derive a more manageable representation of this constraint (\ref%
{normass}). From Fenchel's Equality (Eq. (\ref{FenchelEq})), we have $%
\sum_{y}p_{y}^{n}w_{y}^{n}=\mathcal{G}_{n}(w^{n})+\mathcal{G}^*_{n}(p^{n})=%
\mathcal{G}_{n}^{\ast }(p^{n})$ (with $\mathcal{G}_{n}^{\ast }$ defined as
the convex conjugate function of $\mathcal{G}_{n}$). Moreover, from
Proposition 2, we know that $\mathcal{G}_{n}^{\ast }(p^{n})$ can be
characterized as the optimized dual objective function in (\ref{dual1}).
Hence, we see that the constraint $\mathcal{G}_{n}(w^{n})=0$ is equivalent
to $\int z^{n}(\varepsilon )dQ^{n}(\varepsilon )=0$. We introduce this
latter constraint directly and rewrite the dual program 
\begin{eqnarray}
\inf_{z,w} &&\sum_{y}w_{y}p_{y}^{n}+\int z\left( \varepsilon \right)
dQ^{n}\left( \varepsilon \right)  \label{dualn} \\
s.t.~ &&z(\varepsilon )\geq \iota^{y}\cdot\varepsilon-w_{y},\;\forall
y,\;\forall \varepsilon \\
&&\int z\left( \varepsilon \right) dQ^{n}\left( \varepsilon \right) =0.
\end{eqnarray}%
}

{ We will demonstrate consistency by showing that $(z^{n},w^{n})$
converge a.s. to the dual optimizers in the \textquotedblleft
limit\textquotedblright\ dual problem, given by 
\begin{eqnarray}
\inf_{z,w} &&\sum_{y}w_{y}p_{y}^{0}  \label{dualinf} \\
&&z(\varepsilon )\geq \iota^{y}\cdot\varepsilon -w_{y},\;\forall y,\;\forall
\varepsilon \\
&&\int z\left( \varepsilon \right) dQ=0
\end{eqnarray}%
We denote the optimizers in this limit problem by $(w^{0},z^{0})$, where, by
construction, $w^{0}$ are the \textquotedblleft true\textquotedblright\
values of the choice-specific value functions. The difference between the
discretized and limit dual problems is that $Q^{n}$ in the former has been
replaced by $Q$, the continuous distribution of $\varepsilon $, and the
estimated choice probabilities $p^{n}$ have been replaced by the limit $%
p^{0} $. }

{ We proceed in two steps. First, we argue that the sequence of
optimized dual programs (\ref{dualn}) converges to the optimized limit dual
program (\ref{dualinf}), a.s. Based upon this, we then argue that the
sequence of dual optimizers, $(w^{n},z^{n})$, necessarily converge to their
unique limit optimizers, $(w^{0},z^{0})$, a.s. }

{ {\bfseries First step.} By the Kantorovich duality theorem, we
know that the optimized values for the limit primal and dual programs
coincide 
\begin{equation}
\sup_{\gamma \in \Pi (Q^{0},p^{0})}\int_{{\mathbb{R}}^{d}\times {\mathbb{R}}%
^{d}}\left( \iota\cdot\varepsilon\right) \gamma (d\varepsilon ,d\iota)=\inf
\sum_{y}w_{y}p_{y}^{0}+\int z\left( \varepsilon \right) dQ.  \label{kantor}
\end{equation}
}

{ Moreover, both the primal and dual problems in the discretized
case are finite-dimensional linear programming problem, and by the usual LP
duality, the optimal primal and dual problems for the discretized case also
coincide: 
\begin{equation*}
\int_{{\mathbb{R}}^{d}\times {\mathbb{R}}^{d}}\left( \iota\cdot\varepsilon
\right) \gamma _{n}(d\varepsilon ,d\iota)=\sum_{y}w_{y}^{n}p_{y}^{n}+\int
z^{n}\left( \varepsilon \right) dQ^{n}.
\end{equation*}
}

{ Given Assumption 1, and by Theorem 5.20 in Villani (2009), p.
77, we have that, up to a subsequence extraction, $\gamma ^{n}$ (the
optimizing argument of (\ref{mkn})) converges weakly. 
In addition, by Theorem 5.30 in Villani (2009), the left-hand side of (\ref%
{kantor}) has a unique solution $\gamma $; hence, the sequence $\gamma ^{n}$
must converge generally to $\gamma $. This implies a.s. convergence of the
value of the primal problems: 
\begin{equation*}
\int_{{\mathbb{R}}^{d}\times {\mathbb{R}}^{d}}\left( \iota\cdot\varepsilon
\right) \gamma _{n}(d\varepsilon ,d\iota)\rightarrow \int_{{\mathbb{R}}%
^{d}\times {\mathbb{R}}^{d}}\left( \iota\cdot\varepsilon\right) \gamma
(d\varepsilon ,d\iota),\quad a.s.,
\end{equation*}%
and, by duality, we must also have a.s. convergence of the discretized dual
problem to the limit problem: 
\begin{equation}
\sum_{y}w_{y}^{n}p_{y}^{n}+\int z^{n}\left( \varepsilon \right)
dQ^{n}\rightarrow \sum_{y}w_{y}p_{y}^{0}+\int z\left( \varepsilon \right)
dQ,\quad a.s.  \label{dualconverge}
\end{equation}%
}

{ {\bfseries Second step.} Next, we show that the discretized
dual minimizers $(z^{n},w^{n})$ converge a.s. For convenience, in what
follows we will suppress the qualifier \textquotedblleft
a.s.\textquotedblright\ from all the statements below. Let%
\begin{equation}
\text{\b{w}}^{n}=\min_{y}w_{y}^{n}.  \label{wlowerbar}
\end{equation}%
}

{ From examination of the dual problem (\ref{dualn}), we see that 
$z^{n}$ is the piecewise affine function 
\begin{equation}
z^{n}(\varepsilon )=\max_{y}\{\iota ^{y}\cdot\varepsilon -w_{y}^{n}\},
\label{convlip}
\end{equation}%
thus $z^{n}$ is $M$-Lipschitz with $M:=\max_{y}|\iota ^{y}|=1$. Now observe
that 
\begin{equation}
z^{n}(\varepsilon )+\text{\b{w}}^{n}=\max_{y}\{\iota^{y}\cdot%
\varepsilon-w_{y}^{n}+\text{\b{w}}^{n}\}\leq \max_{y}\{\iota ^{y}\cdot
\varepsilon \}=:\overline{z}(\varepsilon )  \label{upperbound}
\end{equation}%
and, letting $y^{\prime }$ be the argument of the minimum in~(\ref{wlowerbar}%
), 
\begin{equation}
z^{n}(\varepsilon )+\text{\b{w}}^{n}\geq \iota ^{y^{\prime
}}\cdot\varepsilon -w_{y^{\prime }}^{n}+\text{\b{w}}^{n}=\iota ^{y^{\prime
}}\cdot\varepsilon \geq \min_{y}\{\iota ^{y}\cdot \varepsilon \}=:\underline{%
z}(\varepsilon )  \label{lowerbound}
\end{equation}%
thus, by a combination of~(\ref{upperbound}) and~(\ref{lowerbound}), 
\begin{equation}
\underline{z}(\varepsilon )\leq z^{n}(\varepsilon )+\text{\b{w}}^{n}\leq 
\overline{z}(\varepsilon ).  \label{lip}
\end{equation}%
By $\int z^{n}\left( \varepsilon \right) dQ^{n}\left( \varepsilon \right) =0$%
, we have that that \b{w}$^{n}$ is uniformly bounded (sublinear): 
for some constant $K$, $|z^{n}(\varepsilon )|\leq C(1+|\varepsilon |)$ for
every $n$ and every $\varepsilon $. Hence the sequence $z^{n}$ is uniformly
equicontinuous, and converges {locally} uniformly up to a subsequence
extraction by Ascoli's theorem. Let this limit function be denoted $z^{0}$.
By (\ref{dualconverge}), and Theorem 2, we deduce that $z $, the optimizer
in the limit dual problem is unique\footnote{{ Although the
support of $\varepsilon $ is not bounded, the locally uniform convergence of 
$z^{n}$ and the fact that the second moments of $Q^{n}$ are uniformly
bounded are enough to conclude.}}, so that it must coincide with the limit
function $z^{0}$. }

{ By the definition of $(w^n, z^n)$ as optimizing arguments for (%
\ref{dualn}), we have $\sum_{y} w_{y}^{n} p_{y}^{n}\leq \sum_{y} \text{\b{w}}%
^{n} p_{y} + \int [\overline{z}\left( \varepsilon \right) 
]dQ^n\left( \varepsilon \right)$ or 
\begin{equation*}
\sum_{y}\left( w_{y}^{n}-\text{\b{w}}^{n}\right) p_{y}^{n}\leq \int [%
\overline{z}\left( \varepsilon \right) ]dQ^n\left( \varepsilon \right) =%
\mathbb{E}_{Q^n} \overline{z}
\end{equation*}
The second moment restrictions on $Q^n$ (condition (ii) in the theorem)
imply that $\mathbb{E}_{Q^n} \overline{z}(\varepsilon)$ exists and converges
to $\mathbb{E}_{Q} \overline{z}$. Hence, the nonnegative vectors $\left(
w_{y}^{n}-\text{\b{w}}^{n}\right) $ are bounded; accordingly, the vectors $%
\left( w_{y}^{n}\right) $ are themselves bounded. This implies that $w^{n}$
converges up to a subsequence to some limit point $w^{\ast }$, using the
Bolzano-Weierstrass theorem. This implies that $\sum_{y}w_{y}^{n}p_{y}^{n}%
\rightarrow \sum_{y}w_{y}^{\ast }p_{y}$ by bounded convergence. By Theorem
2, we know that the limit point $w^{\ast }$ must coincide with $w^{0}$,
which is the unique optimizer in the dual limit problem (\ref{dualinf}).
Thus, we have shown that $w^n$ converges to $w^0$, a.s. }
\end{proof}

{ \newpage }

\section{\protect\normalsize Additional Figures}

{ 
\begin{table}[h]
{ 
\begin{tabular}{|c|c|c|c|c|}
\hline
Design & RMSE($y=0$) & RMSE($y=1$) & $R^2(y=0)$ & $R^2(y=1)$ \\ \hline
$N=100, T=100$ & 0.5586 (3.7134) & 0.2435 (0.1155) & 0.3438 (0.7298) & 
0.7708 (0.2073) \\ 
$N=100, T=500$ & 0.1070 (0.0541) & 0.1389 (0.0638) & 0.7212 (0.2788) & 
0.9119 (0.0820) \\ 
$N=100, T=1000$ & 0.0810 (0.0376) & 0.1090 (0.0425) & 0.8553 (0.1285) & 
0.9501 (0.0352) \\ \hline
$N=200, T=100$ & 0.1244 (0.0594) & 0.1642 (0.0628) & 0.5773 (0.6875) & 
0.8736 (0.1112) \\ 
$N=200, T=200$ & 0.1177 (0.0736) & 0.1500 (0.0816) & 0.7044 (0.2813) & 
0.9040 (0.0842) \\ \hline
$N=500, T=100$ & 0.0871 (0.0375) & 0.1162 (0.0430) & 0.8109 (0.2468) & 
0.9348 (0.0650) \\ 
$N=500, T=500$ & 0.0665 (0.0261) & 0.0829 (0.0290) & 0.8899 (0.1601) & 
0.9678 (0.0374) \\ \hline
$N=1000, T=100$ & 0.0718 (0.0340) & 0.0928 (0.0344) & 0.8777 (0.1320) & 
0.9647 (0.0314) \\ 
$N=1000, T=1000$ & 0.0543 (0.0176) & 0.0643 (0.0162) & 0.9322 (0.0577) & 
0.9820 (0.0101) \\ \hline
\end{tabular}
}
\caption{}
\label{mc2}
\end{table}
}

{ 
\begin{figure}[]
{ \centering
\includegraphics[scale=0.8]{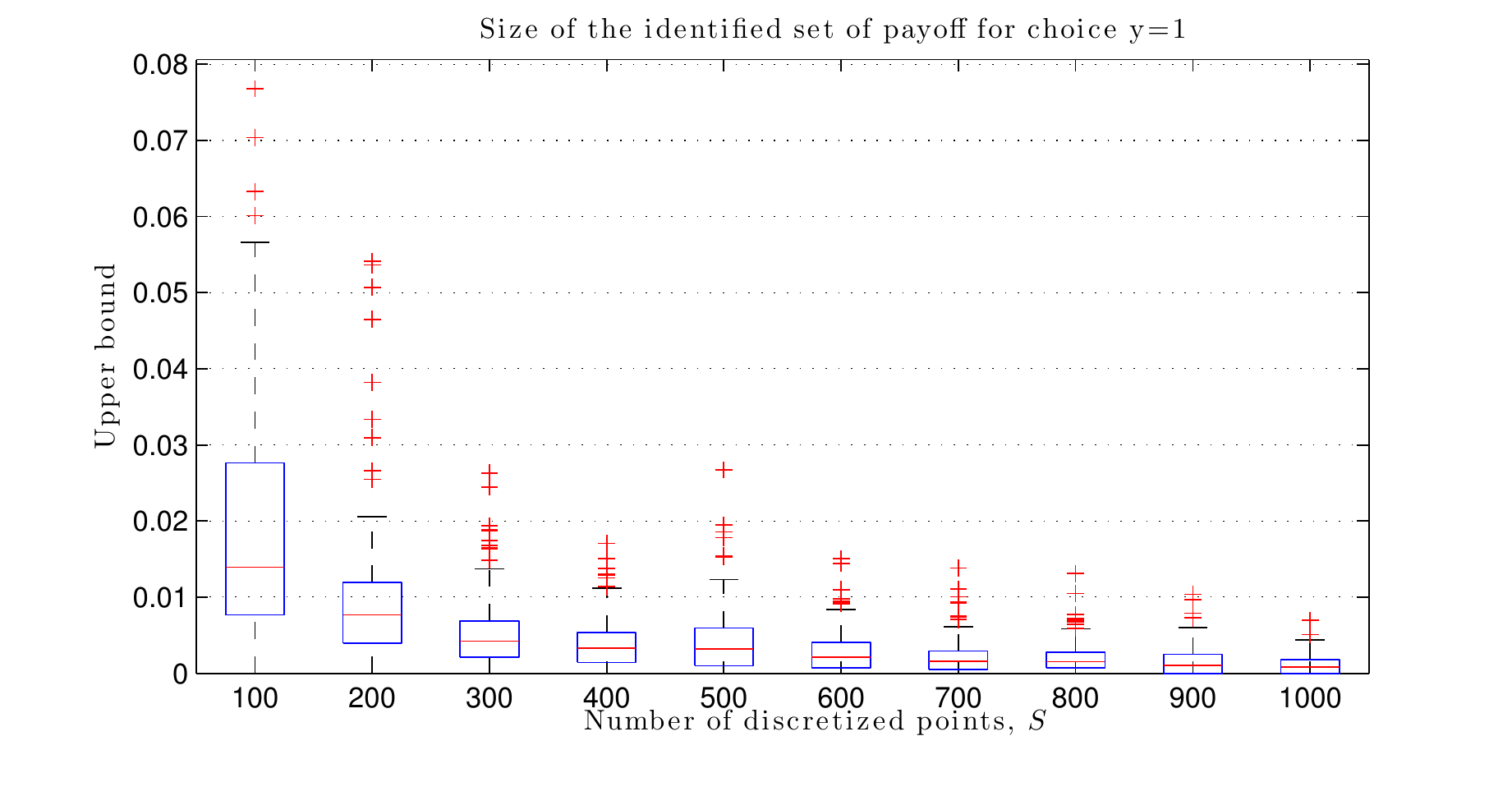}  }
\caption{For each value of $S$, we plot the values of the differences $%
\max_{w\in \partial\mathcal{G}^*(p)} w_1-\min_{w\in \partial\mathcal{G}%
^*(p)} w_1$ across all values of $p\in\Delta^3$. In the boxplot, the central
mark is the median, the edges of the box are the 25th and 75th percentiles,
the whiskers extend to the most extreme data points not considered outliers,
and outliers are plotted individually.}
\label{c1}
\end{figure}
}

{ 
\begin{figure}[tbp]
{ \centering
\includegraphics[scale=0.8]{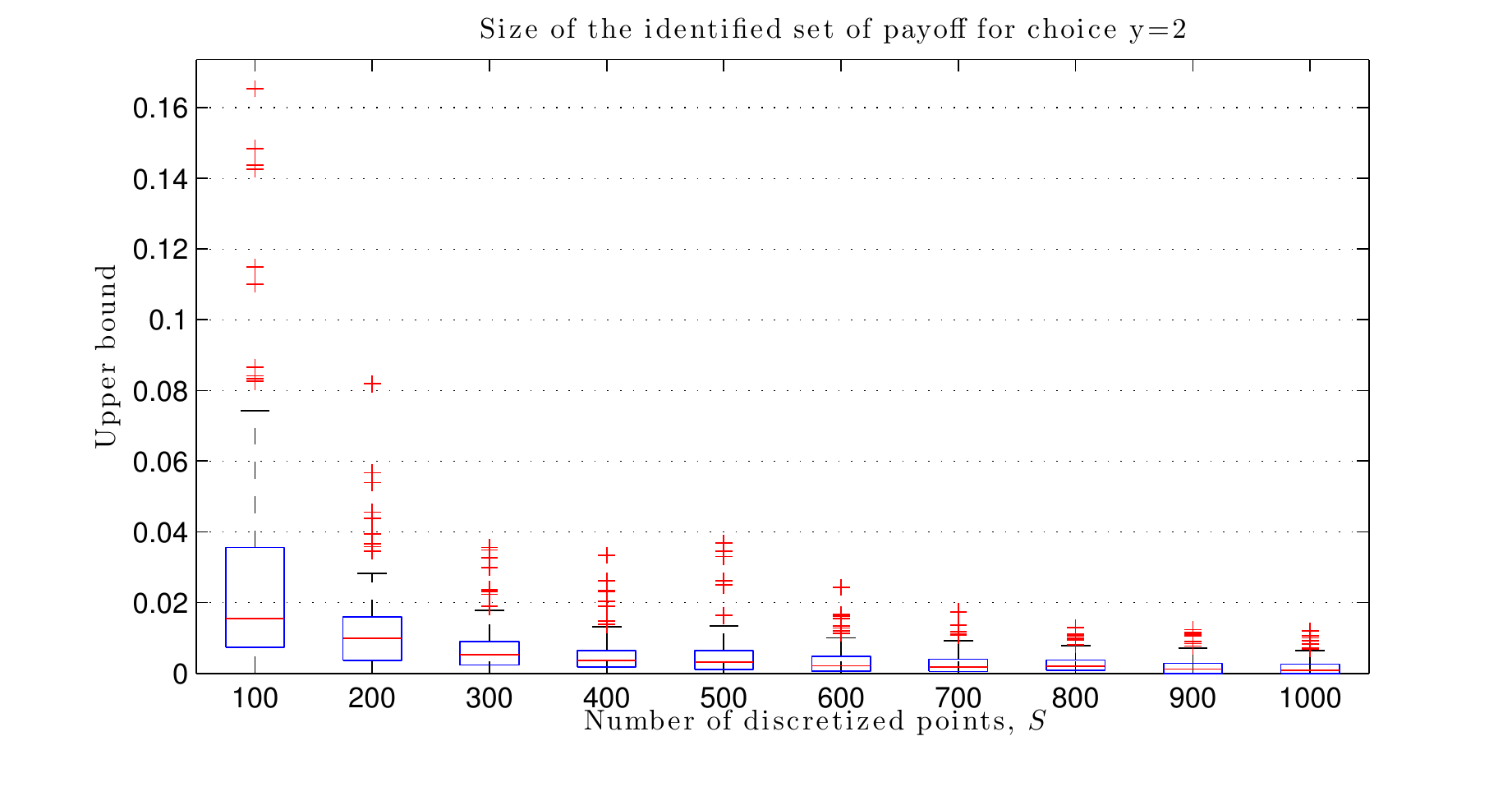}  }
\caption{For each value of $S$, we plot the values of the differences $%
\max_{w\in \partial\mathcal{G}^*(p)} w_2-\min_{w\in \partial\mathcal{G}%
^*(p)} w_2$ across all values of $p\in\Delta^3$. In the boxplot, the central
mark is the median, the edges of the box are the 25th and 75th percentiles,
the whiskers extend to the most extreme data points not considered outliers,
and outliers are plotted individually.}
\label{c2}
\end{figure}
}

{ 
\begin{figure}[tbp]
{ \centering
\includegraphics[scale=0.75]{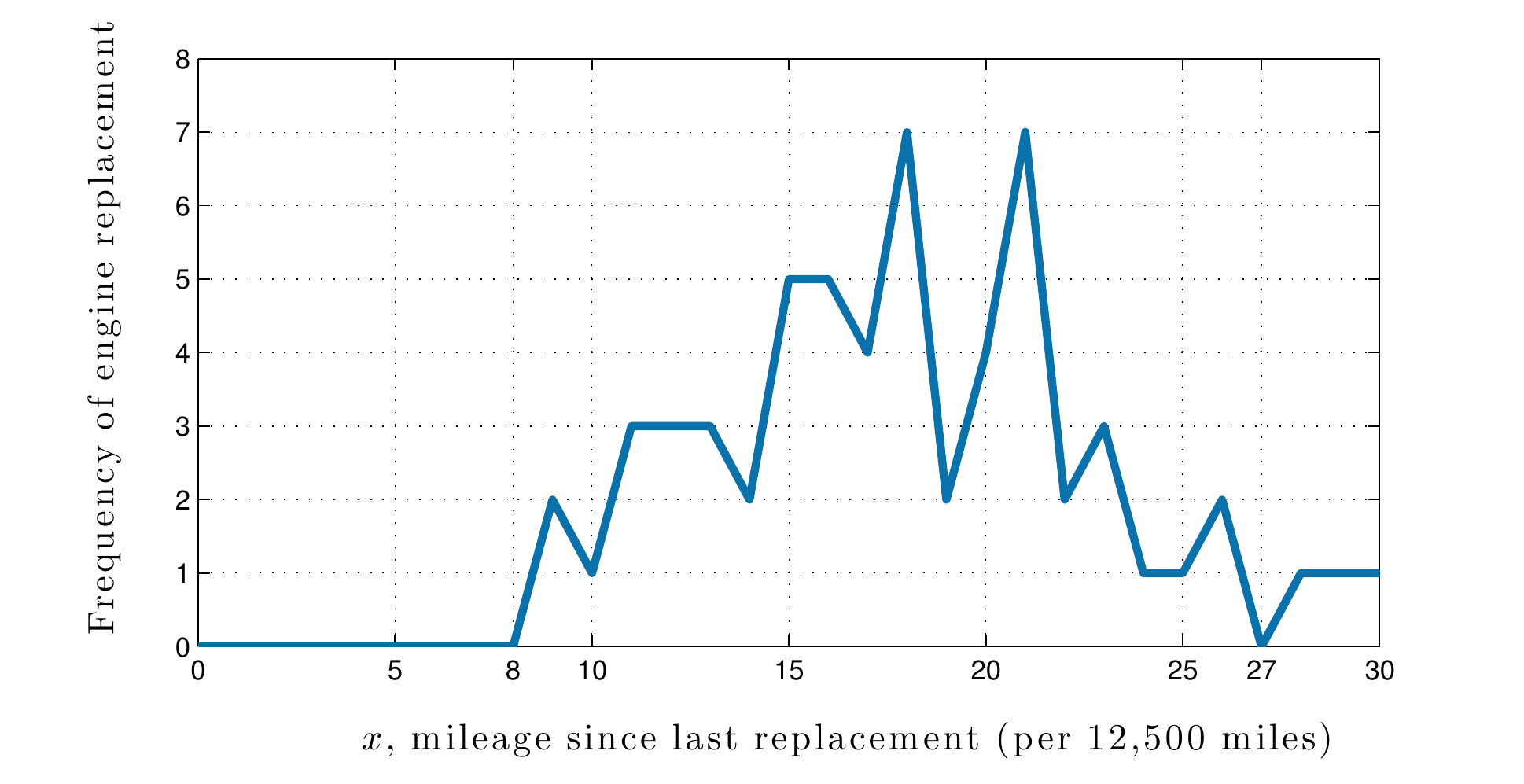}  }
\caption{}
\label{fig:numreplace}
\end{figure}
}

{ 
\begin{figure}[tbp]
{ \centering
\includegraphics[scale=0.75]{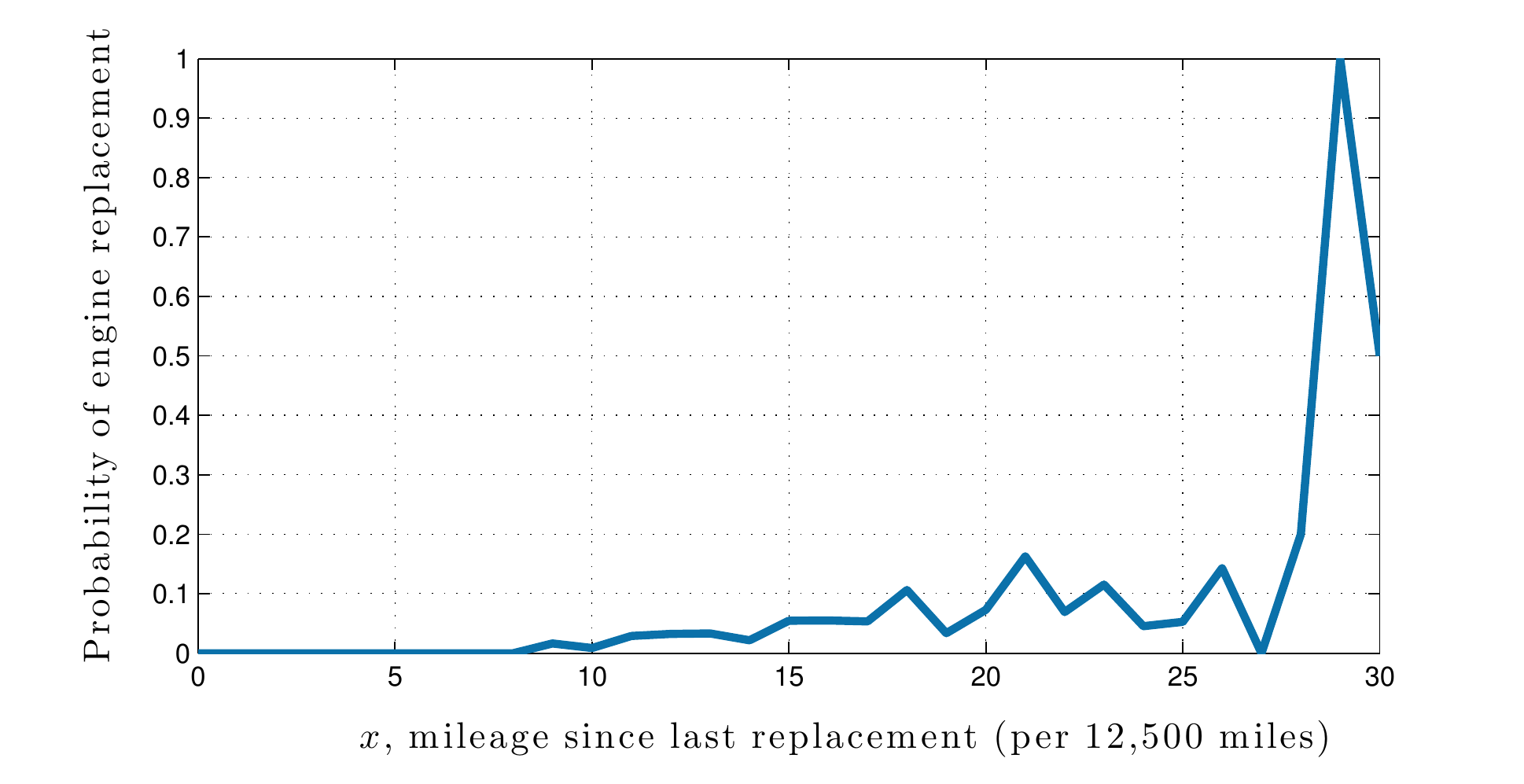}  }
\caption{}
\label{fig:ccp}
\end{figure}
}


\begin{thebibliography}{99}
\bibitem{aguirregabiria_swapping} { V. Aguirregabiria and P.
Mira. Swapping the nested fixed point algorithm: A class of estimators for
discrete Markov decision models. \emph{Econometrica}, 70:1519-1543, 2002. }

\bibitem{aguirregabiria_dynamicgames} { V. Aguirregabiria and P.
Mira. Sequential estimation of dynamic discrete games. \emph{Econometrica},
75:1--53, 2007. }

\bibitem{aguirregabiria_survey} { V. Aguirregabiria and P.
Mira. Dynamic discrete choice structural models: a survey. \emph{Journal of Econometrics},
156:38--67, 2010. }

\bibitem{ADT} { Anderson, S., de Palma, A., and Thisse, J.-F. A
Representative Consumer Theory of the Logit Model. \emph{International
Economic Review}, 29(3), 461-466, 1988. }

\bibitem{ArcidiaconoMiller2011} { P. Arcidiacono and R. Miller.
Conditional Choice Probability Estimation of Dynamic Discrete Choice Models
with Unobserved Heterogeneity. \emph{Econometrica}, 79: 1823-1867, 2011. }

\bibitem{ArcidiaconoMiller2013} { P. Arcidiacono and R. Miller.
Identifying Dynamic Discrete Choice Models off Short Panels. Working paper,
2013. }

{ 
}

\bibitem{hitchhiker} { C. Aliprantis and K. Border. \emph{%
Infinite Dimensional Analysis: A Hitchhiker's Guide}. Springer-Verlag, 2006. 
}

\bibitem{bajari_hong} { P. Bajari, V. Chernozhukov, H. Hong, and
D. Nekipelov. Nonparametric and semiparametric analysis of a dynamic game
model. Preprint, 2009. }

\bibitem{BerryGandhiHaile2012} { S. Berry, A. Gandhi, and P.
Haile. Connected Substitutes and Invertibility of Demand. \emph{Econometrica}
81: 2087-2111, 2013. }

\bibitem{Berry1994} { S. Berry. Estimating Discrete-Choice models
of Production Differentiation. \emph{RAND Journal of Economics}, 25:242-262,
1994. }

\bibitem{blp1} { S. Berry, J. Levinsohn, and A. Pakes. Automobile
prices in market equilibrium. \emph{Econometrica}, 63:841--890, July 1995. }

\bibitem{Bertsekas} { D. Bertsekas. {\em Dynamic Programming Deterministic and Stochastic Models}.  Prentice-Hall, 1987.}

\bibitem{ChiapporiKomunjer2010} { P. Chiappori and I. Komunjer.
On the Nonparametric Identification of Multiple Choice Models. Working
paper, 2010. }

\bibitem{ChiongGalichonShum2014} { K. Chiong, A. Galichon, and M.
Shum. Simulation and Partial Identification in Random Coefficient Discrete
Choice Demand Models. Work in progress, 2014. }

\bibitem{CominettiEtal2010} { R. Cominetti, E. Melo, and S.
Sorin. A payoff-based learning procedure and its application to traffic
games. \emph{Games and Economic Behavior}, 70:71-83, 2010. }

\bibitem{GalichonSalanie2012} { A. Galichon and B. Salani\'{e}.
Cupid's invisible hand: Social surplus and identification in matching
models. Preprint, 2012. }

\bibitem{GretskyOstroyZame1999} { N. Gretsky, J. Ostroy, and W.
Zame. Perfect Competition in the Continuous Assignment Model. \emph{Journal
of Economic Theory}, Vol. 85, pp. 60-118, 1999. }

\bibitem{HaileHortacsuKosenok2008} { P. Haile, A. Hortacsu, and
G. Kosenok. On the Empirical Content of Quantal Response Models. \emph{%
American Economic Review}, 98:180-200, 2008. }

\bibitem{HofbauerSandholm2002} { J. Hofbauer and W. Sandholm. On
the Global Convergence of Stochastic Fictitious Play. \emph{Econometrica},
70: 2265-2294, 2002. }

\bibitem{hotz_miller} { J. Hotz and R. Miller. Conditional choice
probabilties and the estimation of dynamic models. \emph{Review of Economic
Studies}, 60:497--529, 1993. }

\bibitem{hotz_miller2} { J. Hotz, R. Miller, S. Sanders, and J.
Smith. A Simulation Estimator for Dynamic Models of Discrete Choice. \emph{%
Review of Economic Studies}, 61:265-289, 1994. }

\bibitem{hu_shum_dynamic} { Y. Hu and M. Shum. Nonparametric
Identification of Dynamic Models with Unobserved Heterogeneity. \emph{%
Journal of Econometrics}, 171: 32-44, 2012. }

\bibitem{k_s} { H. Kasahara and K. Shimotsu. Nonparametric
Identification of Finite Mixture Models of Dynamic Discrete Choice. \emph{%
Econometrica}, 77: 135--175, 2009. }

\bibitem{KeaneWolpin1997} { M. Keane and K. Wolpin. The career
decisions of young men. \emph{Journal of Political Economy}, 105: 473--522,
1997. }

\bibitem{kennan2006} { J. Kennan. A Note on Discrete
Approximations of Continuous Distributions. Mimeo, University of Wisconsin
at Madison, 2006. }

\bibitem{magnac} { T. Magnac and D. Thesmar. Identifying dynamic
discrete decision processes. \emph{Econometrica}, 70:801--816, 2002. }

\bibitem{mcf1978} { D. McFadden. Modeling the choice of
residential location. In A. Karlquist et. al., editor, \emph{Spatial
Interaction Theory and Residential Location}. North Holland Pub. Co., 1978. }

\bibitem{McFadden1981} { D. McFadden. Economic Models of
Probabilistic Choice. In C. Manski and D. McFadden, editors, \emph{%
Structural Analysis of Discrete Data with Econometric Applications}, 1981. }

\bibitem{norets} { A. Norets. Inference in dynamic discrete
choice models with serially correlated unobserved state variables. \emph{%
Econometrica}, 77: 1665-1682, 2009. }

\bibitem{NoretsTakahashi2013} { A. Norets and S. Takahashi. On
the Surjectivity of the Mapping Between Utilities and Choice Probabilities. 
\emph{Quantitative Economics} 4.1 (2013): 149-155. }

\bibitem{NoretsTang2013} { A. Norets and X. Tang. Semiparametric
Inference in Dynamic Binary Choice Models. Preprint, Princeton University,
2013. }

\bibitem{pakes_patents} { A. Pakes. Patents as options: some
estimates of the value of holding European patent stocks. \emph{Econometrica}%
, 54:1027-1057, 1986. }

\bibitem{pesendorfer_dengler} { M. Pesendorfer and P.
Schmidt-Dengler. Asymptotic least squares estimators for dynamic games. 
\emph{Review of Economic Studies}, 75:901--928, 2008. }

\bibitem{rockafellar1996convex} { R. Tyrell Rockafellar. \emph{%
Convex Analysis}. Princeton University Press, 1970. }

\bibitem{Rust1994} { J. Rust. Structural Estimation of Markov
Decision Processes. \emph{Handbook of Econometrics}, Volume 4 (ed. R. Engle
and D. McFadden). North-Holland, 1994. }

\bibitem{rust_bus} { J. Rust. Optimal replacement of GMC bus
engines: An empirical model of Harold Zurcher. \emph{Econometrica},
55:999--1033, 1987. }

\bibitem{sss} { X. Shi, M. Shum, and W. Song. Estimating
Multinomial Models using Cyclic Monotonicity. Caltech Social Science Working
Paper 1397, 2014. }

\bibitem{ShapleyShubik1971} { L. Shapley and M. Shubik. The
assignment game I: The core. \emph{International Journal of Game Theory},
1(1):111--130, 1971. }

\bibitem{Villani2003} { C. Villani. \emph{Topics in Optimal
Transportation}. Graduate Studies in Mathematics, Vol. 58. American
Mathematical Society, 2003. }

\bibitem{Villani2009} { C. Villani. \emph{Optimal Transport, Old
and New}. Springer, 2009. }
\end{thebibliography}
\end{document}